\documentstyle[aps,preprint,tighten,floats,psfig]{revtex}

\begin{document}

\title{Magnetic polarizability of hadrons from lattice QCD in the background field method}

\author{Frank X. Lee and Leming Zhou}
\address{Center for Nuclear Studies, Department of Physics,
The George Washington University, Washington, DC 20052, USA}
\author{Walter Wilcox}
\address{Department of Physics, Baylor University, Waco, TX 76798, USA}
\author{Joe Christensen}
\address{Physics Department, McMurry University, Abilene, TX 79697, USA}

\maketitle

\begin{abstract}
We present a calculation of hadron magnetic polarizability
using the techniques of lattice QCD. This is carried out by
introducing a uniform external magnetic field on the lattice and
measuring the quadratic part of a hadron's mass shift. 
The calculation is performed on a $24^4$ lattice with standard
Wilson actions at beta=6.0 (spacing $a=0.1$ fm) and pion mass down to about 500 MeV.
Results are obtained for 30 particles covering the
entire baryon octet ($n$, $p$, $\Sigma^0$, $\Sigma^-$, $\Sigma^+$,
$\Xi^-$, $\Xi^0$, $\Lambda$ ) and decuplet ($\Delta^0$,
$\Delta^-$, $\Delta^+$, $\Delta^{++}$, $\Sigma^{*0}$,
$\Sigma^{*-}$, $\Sigma^{*+}$, $\Xi^{*0}$, $\Xi^{*-}$, $\Omega^-$),
plus selected mesons ($\pi^0$, $\pi^+$, $\pi^-$, $K^0$, $K^+$,
${K}^-$, $\rho^0$, $\rho^+$, $\rho^-$, $K^{*0}$, $K^{*+}$, $K^{*-}$). 
The results are compared with available values from experiments
and other theoretical calculations.
\end{abstract}

\section{Introduction}

Electric and magnetic polarizabilities are fundamental
properties of hadrons. They determine the dynamical response (deformation) of a
hadron to external electromagnetic fields, and provide valuable and sensitive
information about the internal strong interaction structure of the hadron as a 
composite particle. In
\cite{Christensen:2004ca}, the results of a lattice calculation of
hadron electric polarizability for neutral hadrons, based on the
methods of \cite{Fiebig:1989en}, has been presented. In this
work, we will focus on the first calculation of hadron magnetic polarizability from
lattice QCD. Our aim is to present the lattice data in sufficient detail in order 
to facilitate comparison with  existing information  in this area.

Traditionally, the symbol $\alpha$ is used to
represent the electric dipole polarizability, and $\beta$ for the
magnetic dipole polarizability. 
A considerable number of experiments have been performed with the aim
of measuring nucleon polarizabilities.  Most of these have
specifically targeted the proton, some attempts have also been
made at measuring the neutron polarizabilities. The experimental
values are listed in Tables \ref{table:proton} and
\ref{table:neutron}.

\begin{table}[htb]
\caption{A list of electric and magnetic polarizabilities in units of $10^{-4}$  
fm$^3$ for the proton
from recent experiments. The method
is elastic Compton scattering off the proton.}
\label{table:proton}
\begin{center}
\begin{tabular}{lll}
\hline
                   &   {$\alpha_p$} & {$\beta_p$} \\
\hline
{\rm Illinois (1991)}\cite{Federspiel:1991yd}
& 10.9$\pm 2.2\pm 1.3$  & 3.3$\mp 2.2 \mp 1.3$
\\
{\rm Mainz (1992)}\cite{Zieger:jq}
& 10.62$^{-1.19-1.03}_{+1.25+1.07}$ & 3.58$^{+1.19+1.03}_{-1.25-1.07}$
\\
{\rm Saskatoon (1993)}\cite{Hallin:1993ft}
& 9.8$\pm 0.4 \pm 1.1$ & 4.4$\mp 0.4 \mp 1.1$
\\
{\rm Saskatoon (1995)}\cite{MacGibbon:1995in}
& 12.5$\pm 0.8 \pm 0.5$ & 2.1$\mp 0.8 \mp 0.5$
\\
{\rm MAMI (2001)}\cite{OlmosdeLeon:zn}
& 11.9$\pm 0.5 \mp 1.3 \pm 0.3$
& 1.2$\pm 0.7 \pm 0.3 \pm 0.4$
\\
\hline
\end{tabular}
\end{center}
\end{table}

Despite some shortcomings, measurements for the proton
polarizabilities in these experiments are in reasonable agreement
with each other. The experimental situation regarding the
polarizabilities of the neutron is still quite unsatisfactory.
This is mainly because a neutron Compton scattering experiment
cannot be directly performed.

\begin{table}[htb]
\caption{A list of electric and magnetic polarizabilities in units of $10^{-4}$  fm$^3$ for neutron
from recent experiments. In the method column, Elastic means elastic (or free) Deuteron
Compton Scattering; Quasifree means Quasifree
Deuteron Compton Scattering.}
\label{table:neutron}
\begin{center}
\begin{tabular}{llll}
\hline
{\rm Ref.}                   & {\rm Method}    & {$\alpha_n$} & {$\beta_n$}  \\
\hline
\cite{Schmiedmayer:1991}     & {\rm Neutron Scattering}
& 12.3$\pm 1.5\pm$2.0 & 3.1$\pm 1.6\pm$2.0     \\
\cite{Kossert:2002jc}        & {\rm Quasifree}
& 12.5$\pm 1.8^{+1.1}_{-0.6} \pm 1.1$
& 2.7$\mp 1.8^{+0.6}_{-1.1} \mp 1.1$
\\
\cite{Kolb:2000ix}           & {\rm Quasifree}
& 7.6 $\sim$ 14.0 & 1.2$\sim$7.6     \\
\cite{Hornidge:1999xs}       & {\rm Elastic}
& 5.5$\pm$2.0 & 10.3$\mp$2.0     \\
\cite{Lundin:2002jy}         & {\rm Elastic}
& 8.8$\pm 2.4 \pm 3.0$ & 6.5 $\mp 2.4 \mp 3.0$     \\
\hline
\end{tabular}
\end{center}
\end{table}

The experimental situation on the nucleon can be approximately summarized as the following:
the electric polarizability is roughly the same for the proton and neutron, with a value of 
around $10$ in units of $10^{-4}$ fm$^3$; 
the magnetic polarizability is roughly the same for the proton and neutron, with a value of 
around $3$ in the same units.

On the theoretical side, nucleon polarizabilities have been most
studied in the framework of chiral perturbation theory 
(ChPT)~\cite{BKM91,BKM92,BKM93,Babusci97,Hemmert,Holstein05}.
Other approaches include quark models~\cite{Capstick92,Lipkin92,YDong05},
chiral soliton models~\cite{Cohen92,Scherer92}. 
For reviews on polarizabilities, see Refs.~\cite{Holstein97,Drechsel97}.

\section{Method}
\label{method}

\subsection{Weak field expansion}
\label{poly}
The approach is based on 
the mass shifts of the particles measured both in the absence and presence of magnetic fields in the QCD vacuum.
For small external magnetic fields, the mass shift (in the unit system of $\hbar=1=c$)
\begin{equation}
\label{eq:es} \Delta m(B) \equiv m(B) - m(0),
\end{equation}
is given by
\begin{equation}
\label{eq:delta} \Delta m(B) = - \vec{\mu} \cdot \vec{B} -
\frac{1}{2} \beta \vec{B}^2.
\end{equation}
By averaging $\Delta m(B)$ over the field, $\vec{B}$, and its
inverse, $-\vec{B}$, we will form
\begin{equation}
\label{eq:delta2} \Delta m(B)_{even} = - \frac{1}{2} \beta
\vec{B}^2.
\end{equation}
After we get the even $\vec{B}$ mass shift from the lattice
simulation, we do least-chi-square fits to the data points with a
polynomial
\begin{equation}
\label{eq:expansion} \Delta m(B)_{even} = c_2 B^2 + c_4 B^4 +
\cdots.
\end{equation}
The magnetic polarizability is then the negative quadratic
coefficient
\begin{equation}
\beta = -2 c_2.
\end{equation}
The quartic and higher terms in Eq.(\ref{eq:expansion}) are
included in order to measure possible numerical contamination.

\subsection{Interpolating fields}

The calculation of hadron masses centers around the time-ordered, two-point correlation
function in the QCD vacuum, projected to zero momentum:
\begin{equation}
G(t)=\sum_{\vec{x}}\langle 0\,|\, T\{\;\eta(x)\, \bar{\eta}(0)\;\}\,|\,0\rangle,
\label{cf2pt}
\end{equation}
where $\eta$ is the interpolating field built from quark fields
with the quantum numbers of the hadron under consideration.
On the quark level, $G(t)$ is evaluated in terms of quark propagators by way of path 
integrals. On the hadronic level, $G(t)$ is a sum of exponentials which has the entire mass 
spectrum of a given channel. The ground state can be extracted by fitting $G(t)$ at large times 
where it dominates. 

We consider a wide variety of particles in this study. For octet baryons, 
the local interpolating fields are
\begin{equation}
\eta^{n}=\epsilon^{abc} (d^{aT}C\gamma_5 u^b)d^c,
\end{equation}
\begin{equation}
\eta^{p}=\epsilon^{abc} (u^{aT}C\gamma_5 d^b)u^c,
\end{equation}
\begin{equation}
\eta^{\Sigma^-}=\epsilon^{abc} (d^{aT}C\gamma_5 s^b)d^c,
\end{equation}
\begin{equation}
\eta^{\Sigma^0}={1\over\sqrt{2}}\epsilon^{abc} \left[
 (u^{aT}C\gamma_5 s^b)d^c
+(d^{aT}C\gamma_5 s^b)u^c \right],
\end{equation}
\begin{equation}
\eta^{\Sigma^+}=\epsilon^{abc} (d^{aT}C\gamma_5 u^b)s^c,
\end{equation}
\begin{equation}
\eta^{\Xi^-}=\epsilon^{abc} (s^{aT}C\gamma_5 d^b)s^c,
\end{equation}
\begin{equation}
\eta^{\Xi^+}=\epsilon^{abc} (s^{aT}C\gamma_5 u^b)s^c,
\end{equation}
\begin{equation}
\eta^{\Lambda^8}={1\over\sqrt{6}}\epsilon^{abc} \left[
2(u^{aT}C\gamma_5 d^b)s^c
+(u^{aT}C\gamma_5 s^b)d^c 
-(d^{aT}C\gamma_5 s^b)u^c \right].
\end{equation}
In the above expressions, $C$ is the charge conjugation operator and the superscript $T$
means transpose. The sum over the color indices is implied and the $\epsilon^{abc}$ ensures 
that constructed states are color-singlet.
For simplicity, the explicit dependence of the quark field operators on space-time, $q(x)$, 
is not written out.
The normalization factors are chosen so that in the limit of SU(3)-flavor symmetry all correlation functions 
simplify to that of the proton.
In addition to the octet lambda $\Lambda^8$, we also consider flavor-singlet lambda $\Lambda^S$,
\begin{equation}
\eta^{\Lambda^S} = \epsilon^{abc} \epsilon^{uds} (u^{aT}C\gamma_5 d^b)s^c.
\end{equation}
Using the transpose of the terms in the parenthesis, it may be written as
\begin{equation}
\eta^{\Lambda^S}=-2\epsilon^{abc} \left[
-(u^{aT}C\gamma_5 d^b)s^c
+(u^{aT}C\gamma_5 s^b)d^c 
-(d^{aT}C\gamma_5 s^b)u^c \right],
\end{equation}
which has a structure similar to $\eta^{\Lambda^8}$ except for the coefficient of the first term.
Since SU(3)-flavor symmetry is broken by the strange quark, it is interesting to study an interpolating 
field that is made up of the terms common to both types,
\begin{equation}
\eta^{\Lambda^C}={1\over\sqrt{2}}\epsilon^{abc} \left[
 (u^{aT}C\gamma_5 s^b)d^c
-(d^{aT}C\gamma_5 s^b)u^c \right].
\end{equation}
Note that since the $u$ and $d$ quarks respond to magnetic fields differently, 
the usual isospin symmetry in $u$ and $d$ quarks is explicitly broken.

For decuplet baryons, the local interpolating fields are
\begin{equation}
\eta_\mu^{\Delta^-}=\epsilon^{abc} (d^{aT}C\gamma_\mu d^b) d^c,
\end{equation}
\begin{equation}
\eta_\mu^{\Delta^0}={1\over\sqrt{3}} \epsilon^{abc} \left[ 
2(d^{aT}C\gamma_\mu u^b) d^c
+(d^{aT}C\gamma_\mu d^b) u^c \right],
\end{equation}
\begin{equation}
\eta_\mu^{\Delta^+}={1\over\sqrt{3}} \epsilon^{abc} \left[ 
2(u^{aT}C\gamma_\mu d^b) u^c
+(u^{aT}C\gamma_\mu u^b) d^c \right],
\end{equation}
\begin{equation}
\eta_\mu^{\Delta^{++}}=\epsilon^{abc} (u^{aT}C\gamma_\mu u^b) u^c,
\end{equation}
\begin{equation}
\eta_\mu^{\Sigma^{*-}}={1\over\sqrt{3}} \epsilon^{abc} \left[ 
2(d^{aT}C\gamma_\mu s^b) d^c
+(d^{aT}C\gamma_\mu d^b) s^c \right],
\end{equation}
\begin{equation}
\eta_\mu^{\Sigma^{*0}}={2\over\sqrt{3}} \epsilon^{abc} \left[ 
 (u^{aT}C\gamma_\mu d^b) s^c
+(d^{aT}C\gamma_\mu s^b) u^c
+(s^{aT}C\gamma_\mu u^b) d^c \right],
\end{equation}
\begin{equation}
\eta_\mu^{\Sigma^{*+}}={1\over\sqrt{3}} \epsilon^{abc} \left[ 
2(u^{aT}C\gamma_\mu s^b) u^c
+(u^{aT}C\gamma_\mu u^b) s^c \right],
\end{equation}
\begin{equation}
\eta_\mu^{\Xi^-}={1\over\sqrt{3}} \epsilon^{abc} \left[ 
2(s^{aT}C\gamma_\mu d^b) s^c
+(s^{aT}C\gamma_\mu s^b) d^c \right],
\end{equation}
\begin{equation}
\eta_\mu^{\Xi^0}={1\over\sqrt{3}} \epsilon^{abc} \left[ 
2(s^{aT}C\gamma_\mu u^b) s^c
+(s^{aT}C\gamma_\mu s^b) u^c \right].
\end{equation}
We use the Lorentz index $\mu=3$ in this calculation.
Note that for baryons the correlation function $G(t)$ is a 4x4 matrix in Dirac space. 
In the Dirichlet boundary condition used in this work,
the sum of the upper diagonal components couple to the positive-parity state, 
while the lower components couple to the negative-parity state. 
So only diagonal elements are computed.

For mesons, we consider the pseudo-scalar states
\begin{equation}
\eta^{\pi^-}(x)=\bar{u}(x)\gamma_5 d(x),
\end{equation}
\begin{equation}
\eta^{\pi^0}(x)={1\over\sqrt{2}}\left[ \bar{u}(x)\gamma_5 u(x) + \bar{d}(x)\gamma_5 d(x) \right],
\end{equation}
\begin{equation}
\eta^{\pi^+}(x)=\bar{d}(x)\gamma_5 u(x),
\end{equation}
\begin{equation}
\eta^{K^-}(x)=\bar{u}(x)\gamma_5 s(x),
\end{equation}
\begin{equation}
\eta^{K^0}(x)=\bar{d}(x)\gamma_5 s(x),
\end{equation}
\begin{equation}
\eta^{K^+}(x)=\bar{s}(x)\gamma_5 u(x),
\end{equation}
and the vector states
\begin{equation}
\eta_\mu^{\rho^-}(x)=\bar{u}(x)\gamma_\mu d(x),
\end{equation}
\begin{equation}
\eta_\mu^{\rho^0}(x)={1\over\sqrt{2}}\left[ \bar{u}(x)\gamma_\mu u(x) + \bar{d}(x)\gamma_\mu d(x) \right],
\end{equation}
\begin{equation}
\eta_\mu^{\rho^+}(x)=\bar{d}(x)\gamma_\mu u(x),
\end{equation}
\begin{equation}
\eta_\mu^{K^{*-}}(x)=\bar{s}(x)\gamma_\mu d(x),
\end{equation}
\begin{equation}
\eta^{K^{*0}}(x)=\bar{d}(x)\gamma_\mu s(x),
\end{equation}
\begin{equation}
\eta_\mu^{K^{*+}}(x)=\bar{s}(x)\gamma_\mu u(x).
\end{equation}
Here $\bar{q}=q^+\gamma_0$. For the vector mesons, we average over the spatial Lorentz 
indices $\mu=1,2,3$. Note that in forming the correlation function for the $\pi^0$  state,
there is a term from the local contraction in the form of
\begin{equation}
{1\over\sqrt{2}}\left[M^{-1}_u(x,x)-M^{-1}_d(x,x)\right]
\end{equation}
where $M^{-1}_q(x,x)$ denotes all-to-all quark propagators (disconnected quark loops).
In the zero magnetic field and SU(2) isospin symmetry in $u$ and $d$ quarks, the term vanishes.
In the presence of magnetic field, however, it is no longer so. 
The same is true for the $\rho^0$  state.
The calculation of these disconnected loops are prohibitively expensive and 
is prone to large statistical errors.
In this work we will ignore the effect of the disconnected loops for these mesons.

\subsection{Lattice Techniques}

We briefly discuss how the magnetic field is
introduced on the lattice. The procedure we use is very
similar to but not exactly the same as the one presented in \cite{Fiebig:1989en}.

In the continuum case, the fermion action is modified
by the minimal coupling prescription
\begin{equation}
\partial_\mu \rightarrow \partial_\mu + i q A_\mu,
\end{equation}
where $q$ is the charge of the fermion field and $A_\mu$ is the
vector potential describing the external field. On the lattice,
the prescription amounts to supplying a phase factor, $e^{i a q
A_\mu}$, to the link variables in a given direction. Choosing
\begin{equation}
A_2 = B x_1 \ \  {\rm \mbox \rm and} \ \  A_0 = A_1 = A_3=0
\end{equation}
a constant magnetic field
\begin{equation}
F_{12} = \partial_1 A_2 - \partial_2 A_1 =B_3=B
\end{equation}
can be introduced in the $z$-direction.

We introduce two dimensionless parameters which characterize the
field. One is given by
\begin{equation} \label{parameter}
\eta = q B a^2.
\end{equation}
The other is the integer lattice length,
\begin{equation}
\rho = x_1/a.
\end{equation}
In terms of these two parameters the phase factor becomes:
\begin{equation}
e^{i a q A_2} = e^{i \eta \rho} \rightarrow (1 + i \eta \rho ),
\end{equation}
where we have linearized the phase factor to mimic the continuum
prescription. The field strength we use will be chosen small
enough to satisfy the linearization requirement\cite{Fiebig:1989en}.

To summarize, the method to place the external magnetic field on
the lattice in the $z$-direction is to multiply each gauge field link variable
in the $y$-direction with a $x$-dependent factor:
\begin{equation} \label{prescription}
U_2 (x) \rightarrow (1 + i \eta \rho ) U_2 (x)
\end{equation}
%

\subsection{Simulation Details}

Most of our results are based on the standard Wilson quark
action on a quenched 24$^4$ lattice with
lattice spacing $a=0.1$ fm. The lattice coupling $\beta = 6.0$.
We have analyzed 150 configurations
to extract hadron magnetic polarizabilities.
Fermion propagators $M^{-1}$ were constructed at six different quark
masses, which correspond to six the $\kappa$ values,
\begin{equation}
\kappa= 0.1515, 0.1525, 0.1535, 0.1540, 0.1545, 0.1555.
\end{equation}
The critical kappa value is $\kappa_c =0.157096$. Using
\begin{equation}
m_q =\frac{1}{2a} (\frac{1}{\kappa} -\frac{1}{\kappa_c})
\end{equation}
we can get the corresponding quark masses are:
230 MeV, 189 MeV, 147 MeV, 126 MeV, 105 MeV, 64 MeV.
The corresponding pion masses are:
1000 MeV, 895 MeV, 782 MeV, 721 MeV, 657 MeV, 512 MeV.
The strange quark mass is set at $\kappa=0.1535$.
The lattice source is ($x, y, z, t$)= (12, 1, 1, 2).

In an exploratory study, the same calculation was performed 
using a different action, namely, the tree-level tadpole-improved
L\"uscher-Weisz gauge action with
lattice spacing $a = 0.17$ fm (or $1/a = 1159$ MeV) set from the
string tension, and the tadpole-improved clover quark action. 
We accumulated 15 configurations and the preliminary results have
been presented in Ref.\cite{Zhou:2002km}.
We will focus on the Wilson results.
The clover results will be shown together with the Wilson results mainly for 
comparison purposes and as an independent check. 
In most cases, the results from the two action schemes are consistent with 
each other with errors, giving us confidence in the calculation.

We used six different values of the parameter in units of $10^{-3}$
\begin{equation}
\eta = 0.0, +0.36, -0.72, +1.44, -2.88, +5.76.
\end{equation}
The $\eta$ values in this sequence are related by a factor of $-2$.
Thus we are able to study the response of a hadron composed of both
up and down (strange) quarks, whose charges are related by the same factor, 
to four different nonzero magnetic fields.
In units of $10^{-3} e^{-1} a^{-2}$, the magnetic field takes the values
-1.08, 2.16, -4.32,  and 8.64.
In physical units, the magnitude of the weakest magnetic field is
2.46 $\times 10^{13} $ tesla.
This is a very strong magnetic field. On the other hand, in the sense
of the mass shift, this is really not that strong. We can estimate the ratio
of mass shift of proton to the mass of proton:
\begin{equation}
\label{eq:ratioofmassshift}
\frac{\delta m}{m} =  \frac{\frac{1}{2}\beta_p B^2}{m}
= 1.60 \times 10^{-5}.
\end{equation}
Here we have used 2$\times 10^{-4}$ fm$^3$ as the value of $\beta_p$.
From this rough estimation we can see that the mass shift
is very small even in such a strong magnetic field.

\section{Results}

In this section, we present the numerical results of our lattice study. 
We calculated the magnetic polarizability for 30
charged and neutral hadrons from their quadratic response to the field. 
Since this is a first calculation of these quantities,
we will show effective-mass plots for the computed mass shifts for 
all of the particles considered at one value (the weakest) of the magnetic field.
They form the basis of our analysis and give an  unbiased view of the quality of our data.
The mass shifts are fitted to the polynomial form discussed in section~\ref{poly} from which the 
magnetic polarizability is extracted.
We group the results according to the particle types: octet, decuplet, and mesons.

\subsection{Octet Baryons}

\begin{figure}
\centerline{\psfig{file=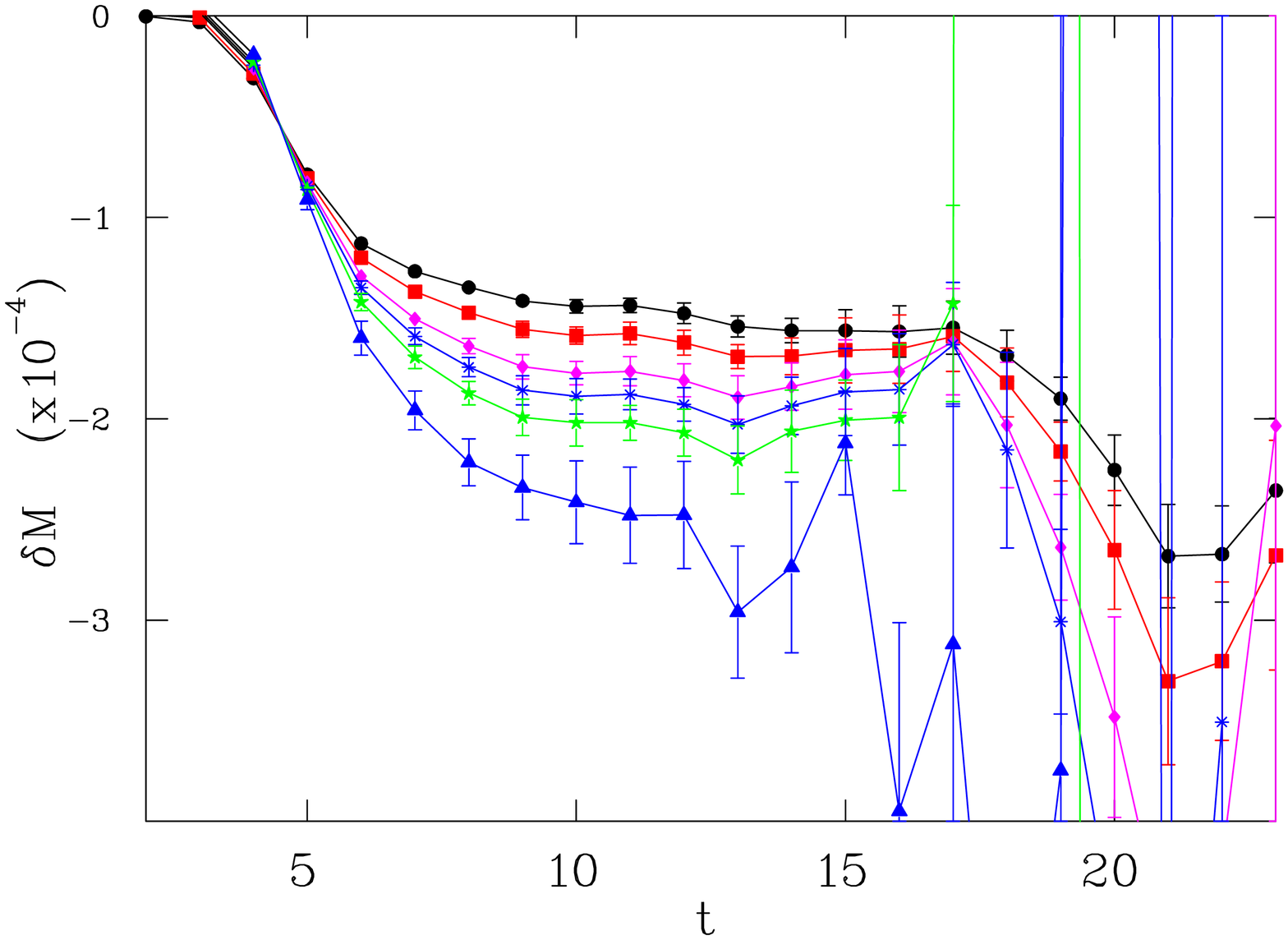,width=11cm,angle=90}}
\centerline{\psfig{file=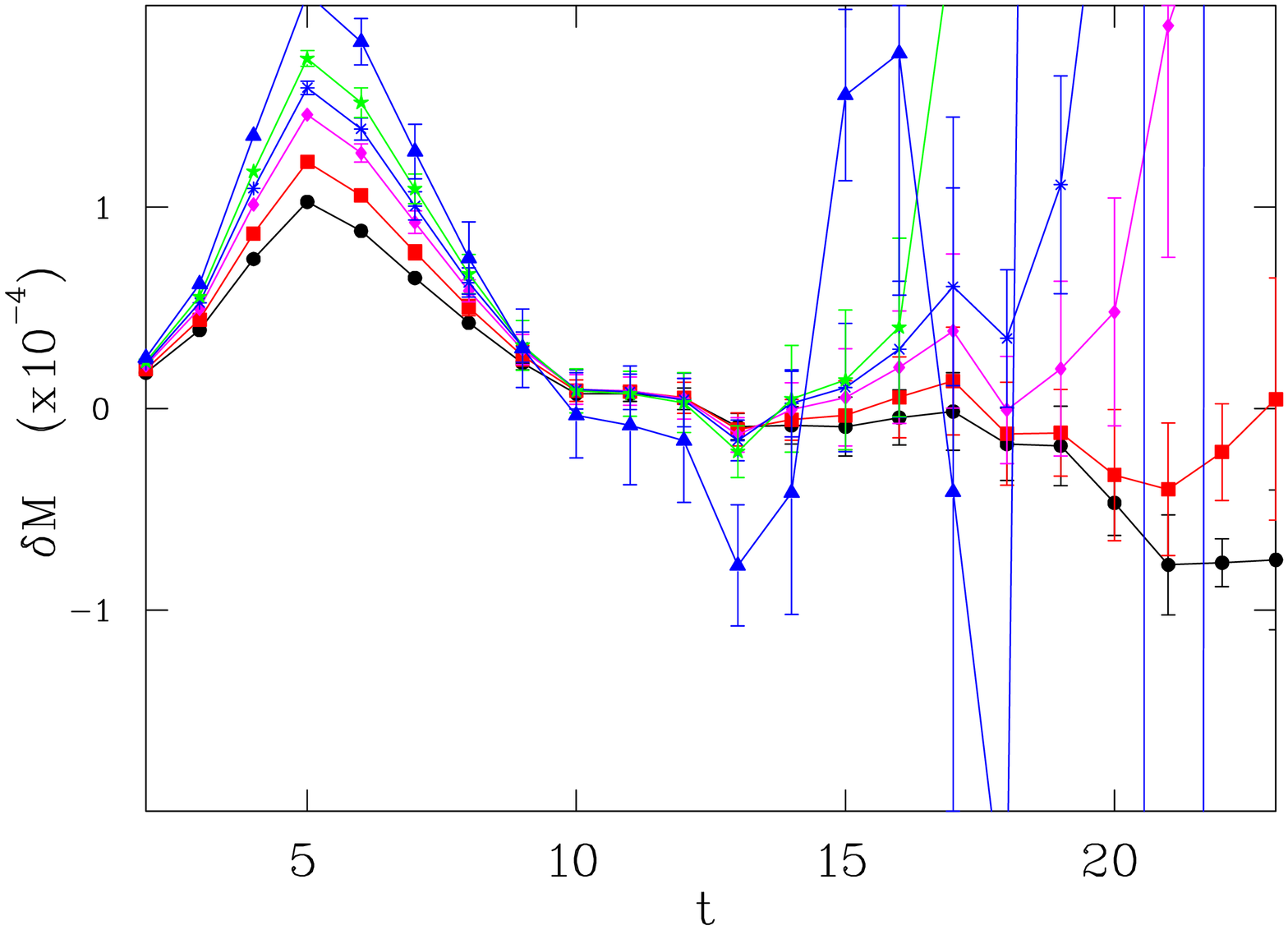,width=11cm,angle=90}}
 \vspace*{+0.5cm}
\caption{Effective mass plots for the neutron (upper) and proton (lower) mass shifts 
in lattice units at the weakest magnetic field.
The lines correspond to quark masses from the heaviest (circles) to the lightest (triangles).}
\label{emass-np-m1}
\end{figure}

\begin{figure}
\centerline{\psfig{file=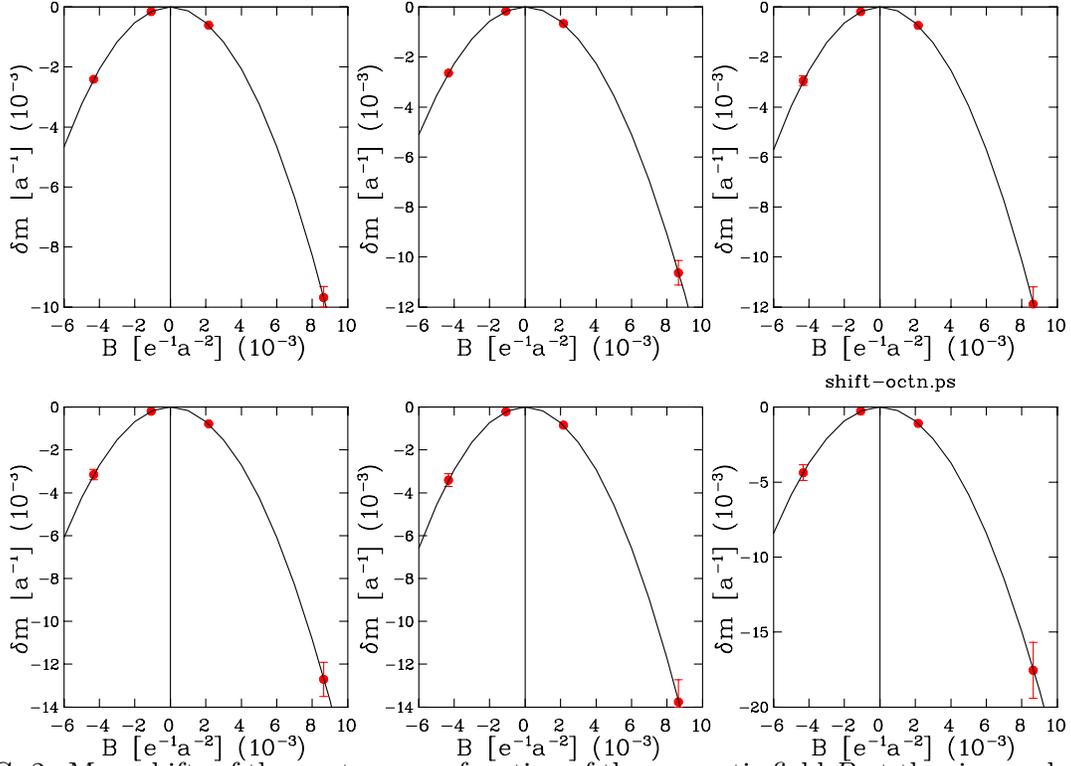,width=14cm,angle=90}}
\caption{Mass shifts of the neutron as a function of the magnetic field 
$B$ at the six quark masses (heavy to light in the order of top left-to-right 
to bottom left-to-right).}
\label{shift-octn}
\end{figure}

Fig.~\ref{emass-np-m1} shows the effective mass plot for the neutron and proton.
There is reasonable plateau behavior for the neutron and our results are extracted from the time 
window of 12 to 14.
The plateau behavior for the proton is not as good as for the neutron.
Fig.~\ref{shift-octn} shows the mass shifts for the neutron as a function of 
the magnetic fields.
There is good parabolic behavior going through the origin, 
an indication that contamination from the linear term has been 
effectively eliminated by averaging results from $\vec{B}$ and $-\vec{B}$.
It is typical of all the particles so we will not show more such plots.
We tried a number of ways in fitting to the data. First, we did the fitting with the $B^2$ term only, 
or with the $B^2$ term plus the $B^4$ term. We found that the $B^4$ term is numerically 
small. This confirms that the magnetic fields we use are indeed weak.
We also tried fits using only the two smallest fields in magnitude, or the three smallest, 
or all four values.  They all give results consistent within errors bars. 
The results quoted below are mostly from fitting to the two smallest fields.

For charged particles, there is the possibility of Landau levels on the order of 
$|qB/(2M)|$ in the presence of magnetic fields, where $q$ and $M$ are the charge and 
mass of the particle, respectively. It is a linear term that is not eliminated by the averaging 
procedure. Their effects only show up at very large times, larger than where we fit the data.
We studied this issue by including a linear term in the fit and found that the results 
for $c_2$ are essentially unaffected. 

\begin{figure}
\centerline{\psfig{file=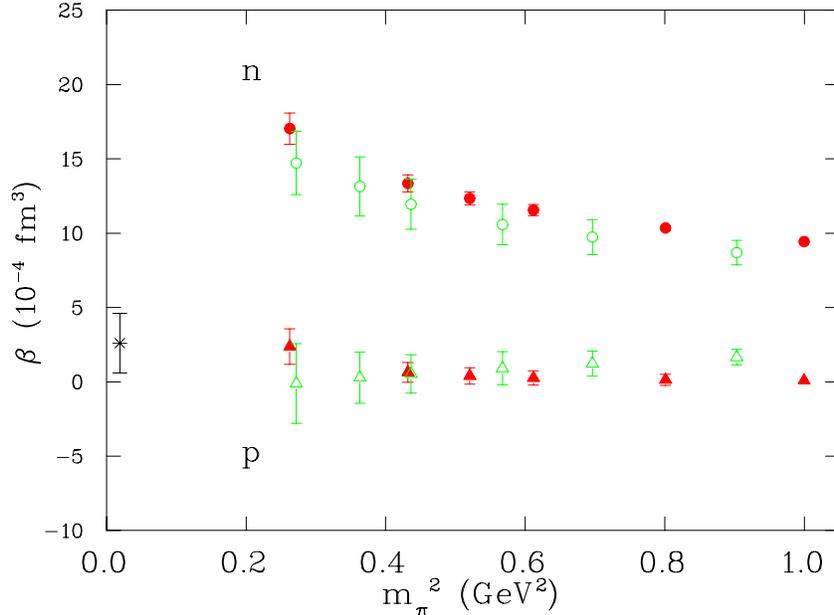,width=11cm,angle=90}}
\caption{Magnetic polarizability of the neutron and proton as a function 
of $m_\pi^2$ in physical units. The solid symbols are the results from the Wilson 
action, while the empty symbols are from the clover action. 
The Wilson results are obtained from the time window of 12 to 14. 
The experimental value, 
which we take to be the same for neutron and proton, is indicated by the star.}
\label{mpol-np-wc}
\end{figure}

Fig.~\ref{mpol-np-wc} shows the extracted magnetic polarizability 
for the neutron and the proton as a function of pion mass 
squared~\footnote{The pion mass squared is proportional to the quark mass in QCD.} 
in physical units.
Two sets of data are displayed, one from the Wilson action based on 150 configurations, 
one from the clover action based on 15 configurations. The two sets are consistent 
within errors. The experimental value, which is roughly the same for neutron and proton with a
large uncertainty, is also indicated. One can see that the proton results are consistent with 
the experimental value, while the neutron results, which have smaller errors than the proton, 
are much bigger. This difference between the proton and the neutron is one of the surprises of our calculation. 
We do not attempt a chiral extrapolation in this work because the theoretical 
ansatz from chiral effective theories is quite limited at the moment and the systematic uncertainties of the lattice results are not quantified.
The reader can get an idea on the chiral behavior by looking at the quark mass 
dependence in the figures and tables.
The observations we make here and below are based 
on the value at the smallest pion mass and the trend as a function of the quark mass, therefore should be taken as qualitative.
For example, the proton magnetic polarizability has a weaker quark mass dependence than the neutron. At the chiral limit, the agreement with experiment for the proton is expected to remain, and 
the difference between the proton and the neutron is expected to grow.

\begin{figure}
\centerline{\psfig{file=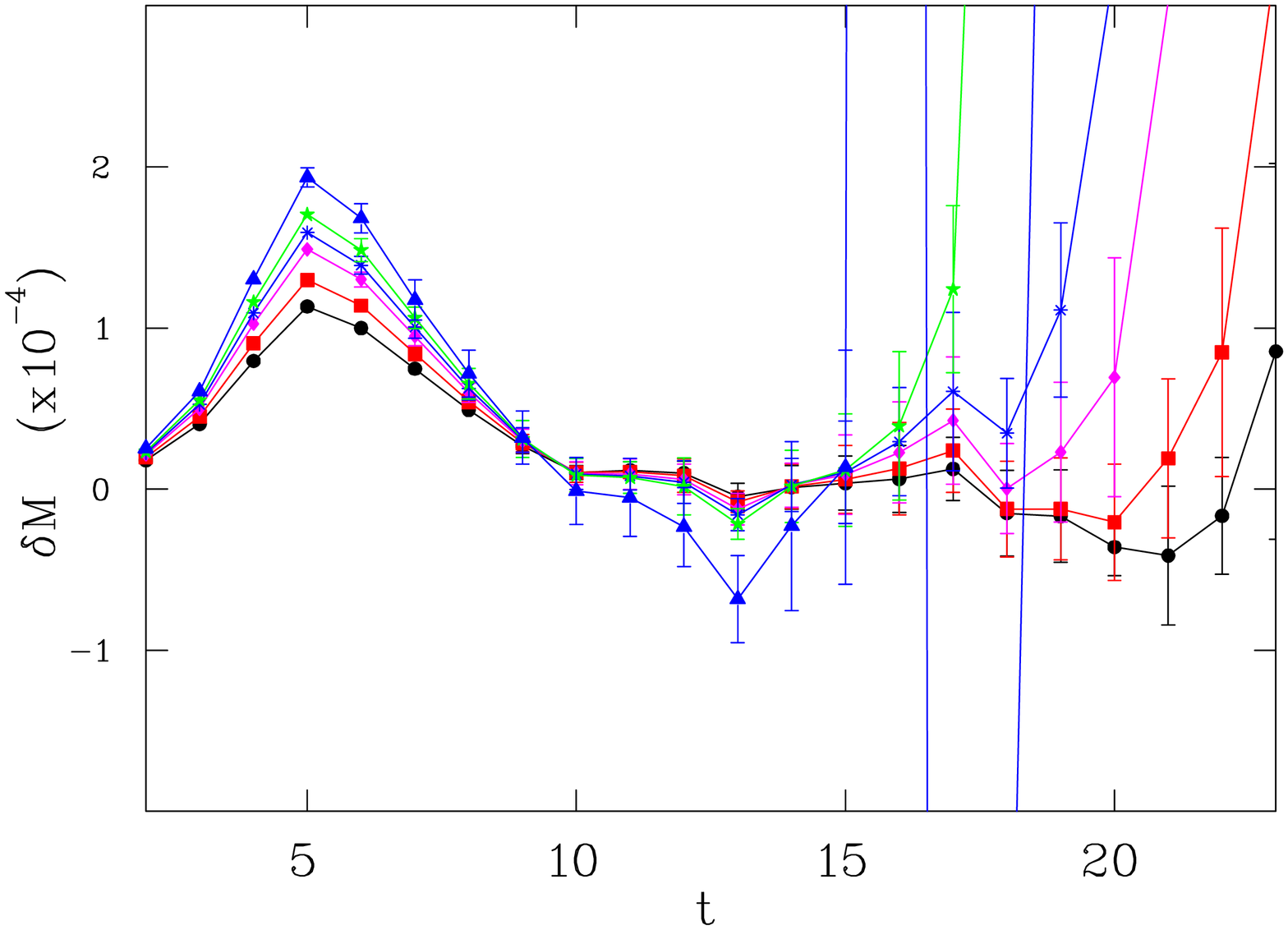,width=9cm,angle=90}}
\centerline{\psfig{file=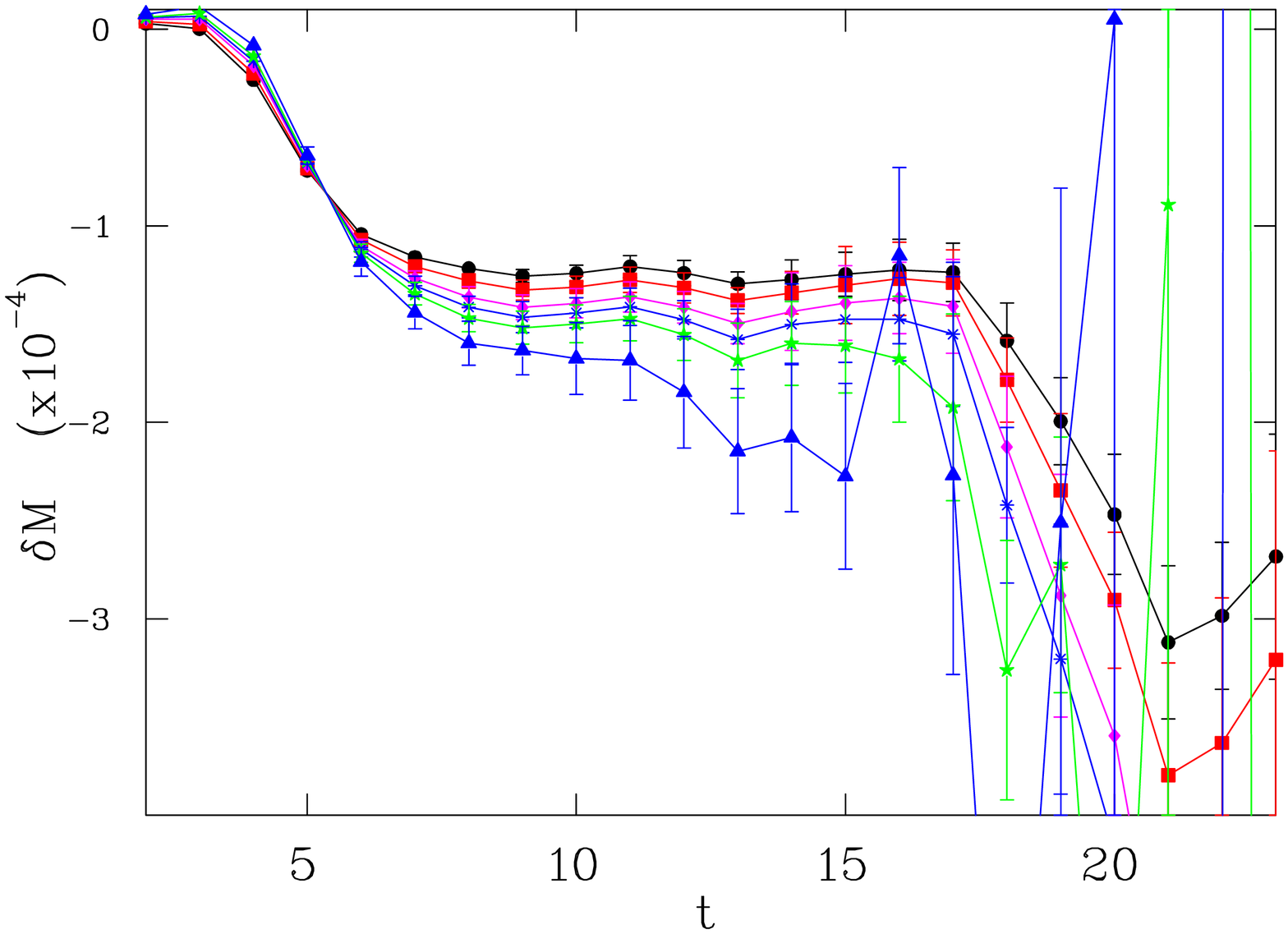,width=9cm,angle=90}}
\centerline{\psfig{file=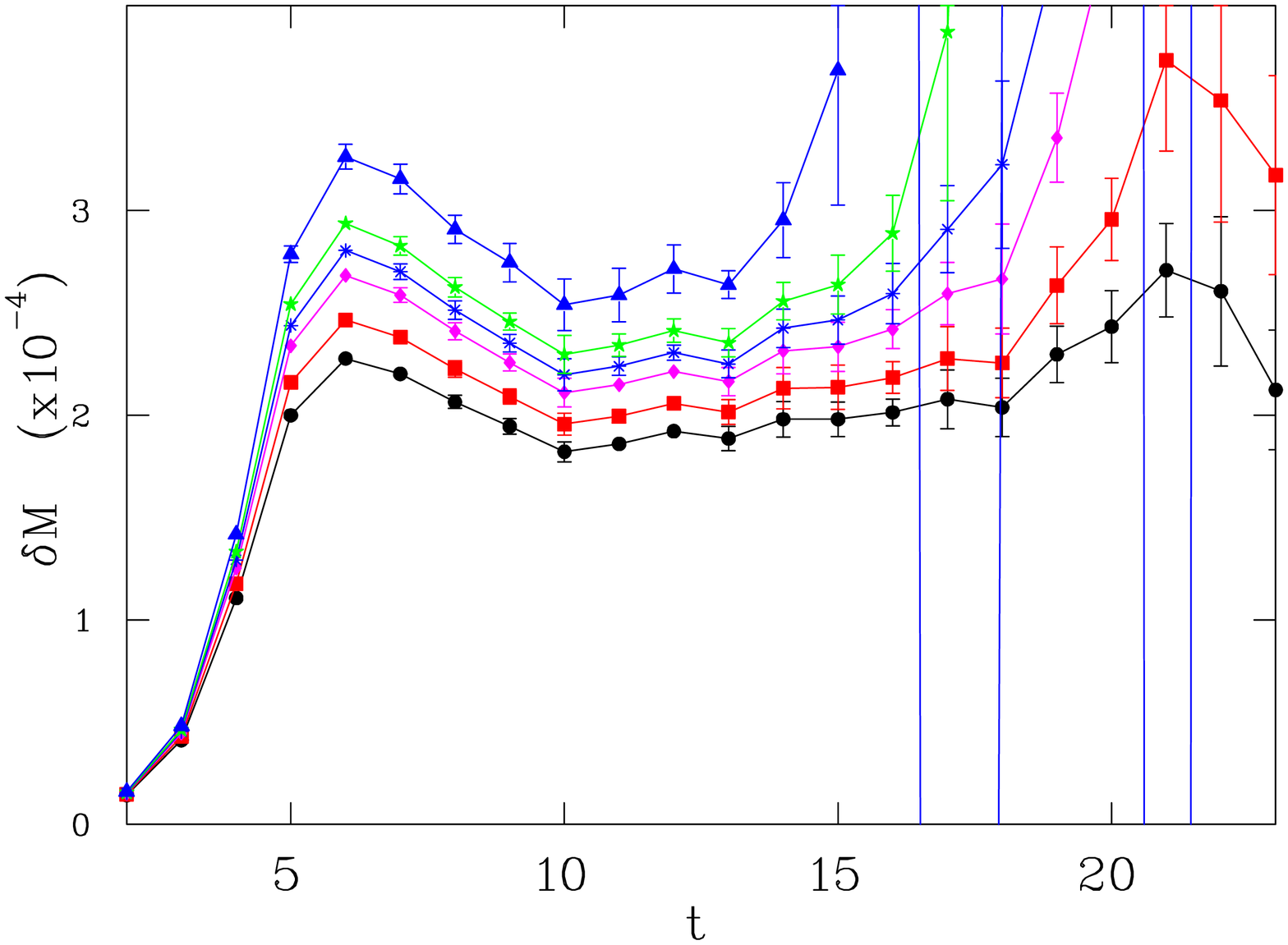,width=9cm,angle=90}}
\caption{Effective mass plots for the octet sigma mass shifts in lattice units 
at the weakest magnetic field 
in the order of $\Sigma^+$ (top), $\Sigma^0$ (middle), $\Sigma^-$ (bottom).
The lines correspond to quark masses from the heaviest (circles) to the lightest (triangles).}
\label{emass-osig-m1}
\end{figure}

\begin{figure}
\centerline{\psfig{file=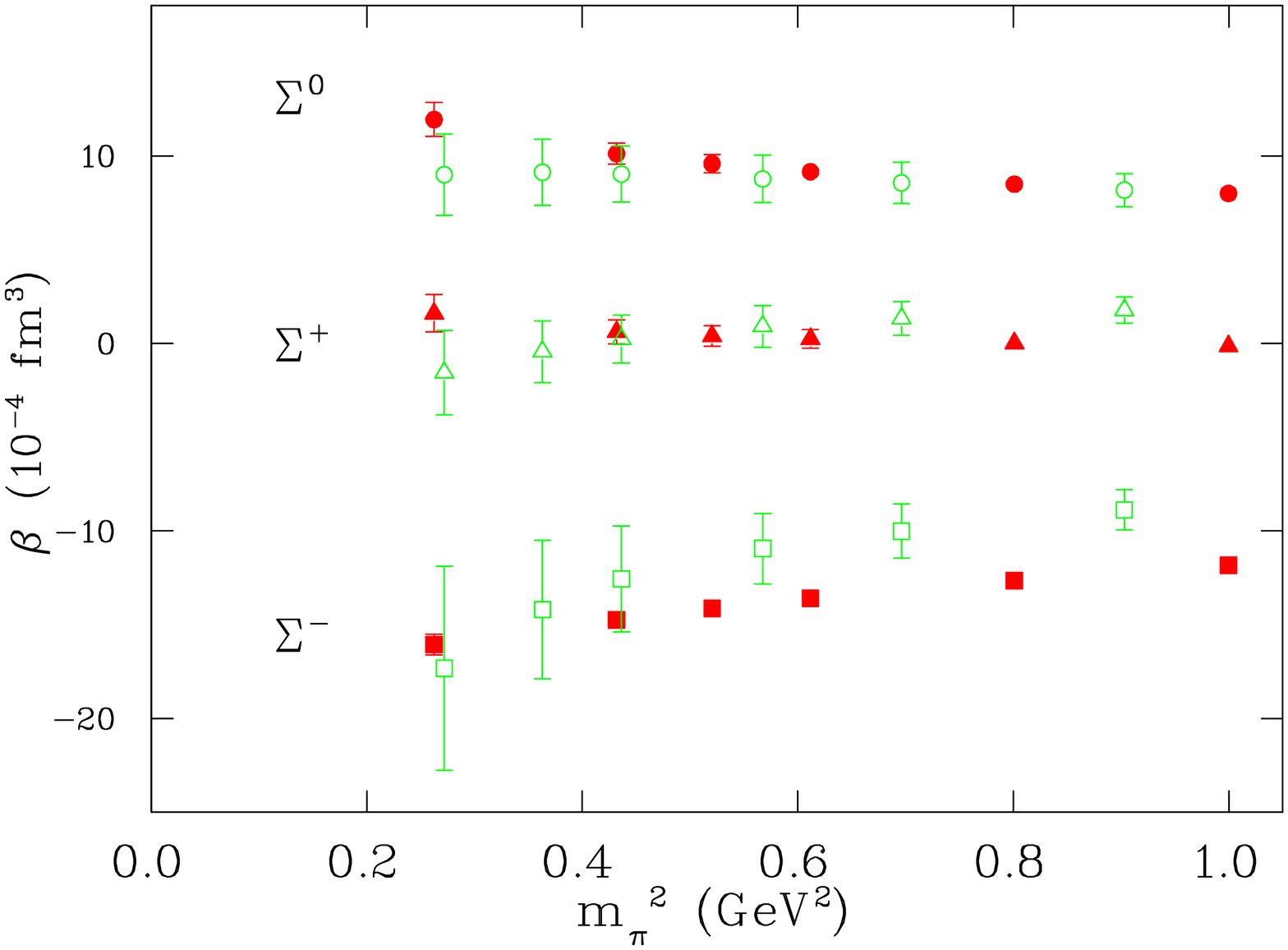,width=11cm,angle=90}}
\caption{Magnetic polarizability of the octet sigma states in physical units. 
The solid symbols are the results from the Wilson 
action, while the empty symbols are from the clover action.
The Wilson results are obtained from the time window of 12 to 14. }
\label{mpol-osig-wc}
\end{figure}

Fig.~\ref{emass-osig-m1} shows the effective mass plot for the octet sigma states.
There is reasonable plateau behavior and we fit in the range of 12 to 14.
Fig.~\ref{mpol-osig-wc} shows the magnetic polarizability for the three octet sigma states. 
The $\Sigma^+$ results are similar to those for the proton and also suffer from 
large errors.  The $\Sigma^0$ results are positive and the $\Sigma^-$ are negative, both have much smaller errors compared to the $\Sigma^+$. 

\begin{figure}
\centerline{\psfig{file=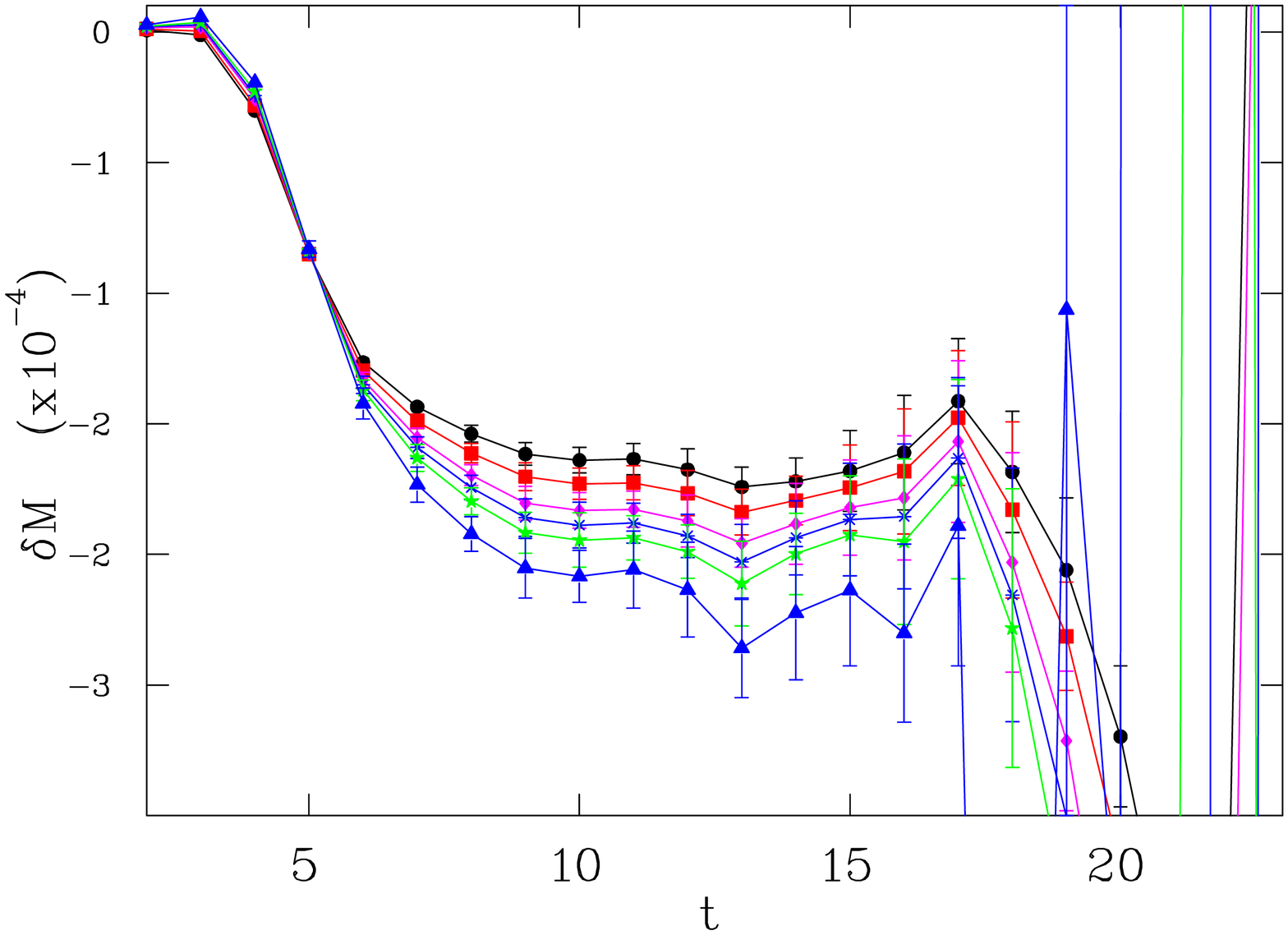,width=7.5cm,angle=90}}
\centerline{\psfig{file=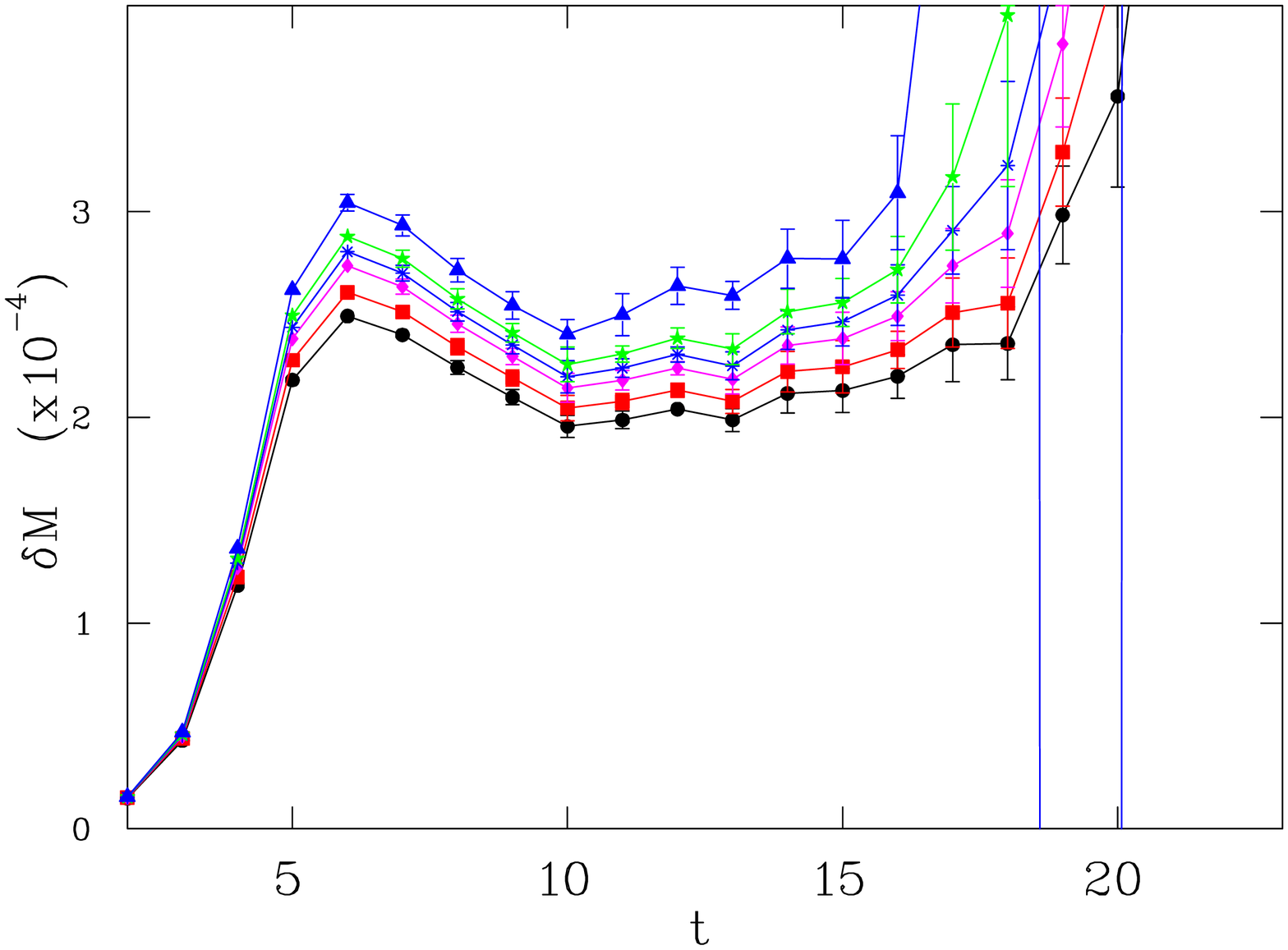,width=7.5cm,angle=90}}
\caption{Effective mass plots for the octet cascade mass shifts in lattice units 
at the weakest magnetic field 
in the order of $\Xi^0$ (upper) and $\Xi^-$ (lower).
The lines correspond to quark masses from the heaviest (circles) to the lightest (triangles).}
\label{emass-oxi-m1}
\end{figure}

\begin{figure}
\centerline{\psfig{file=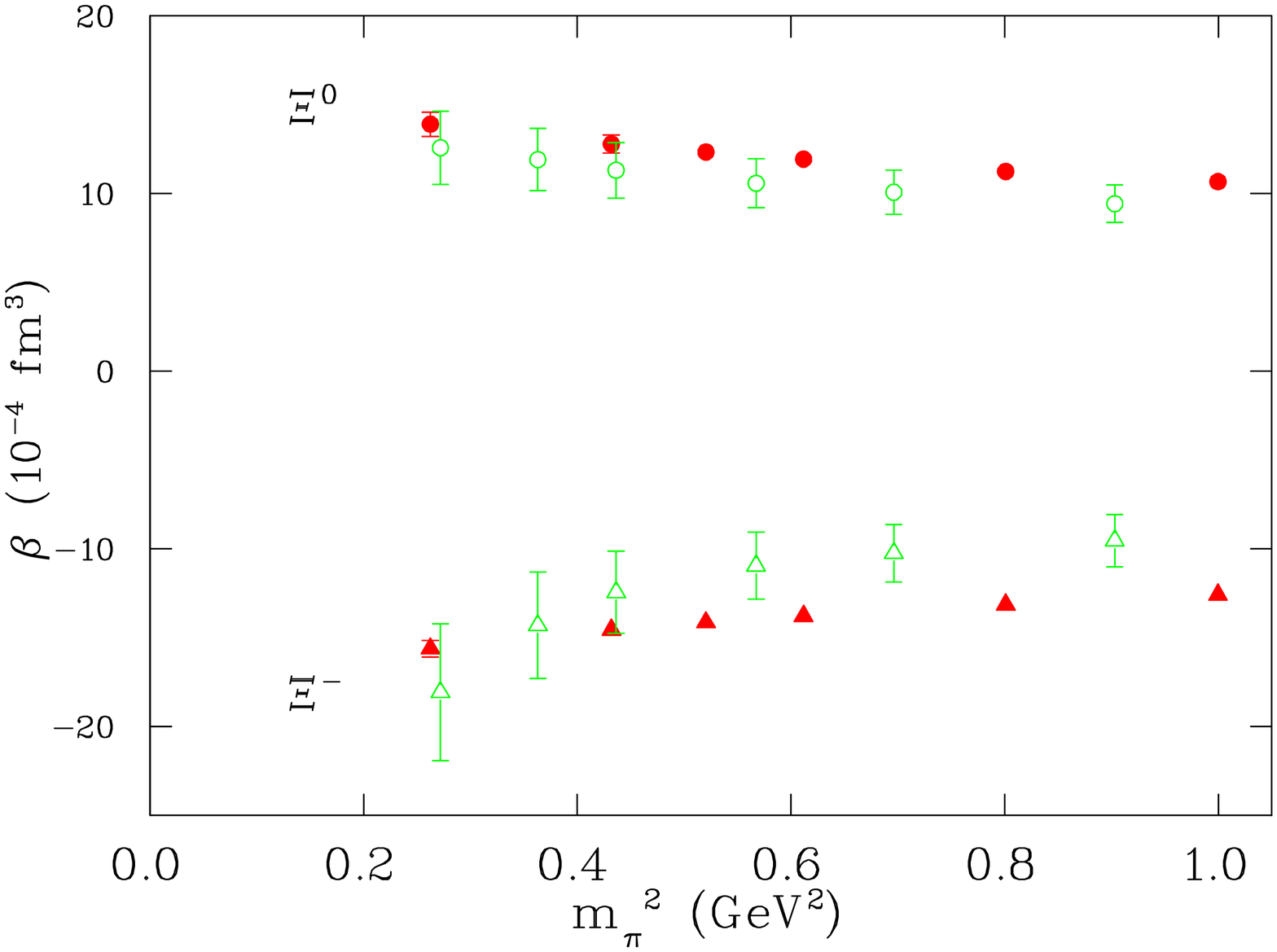,width=11cm,angle=90}}
\caption{Magnetic polarizability of the octet cascade states in physical units. 
The solid symbols are the results from the Wilson 
action, while the empty symbols are from the clover action.
The Wilson results are obtained from the time window of 12 to 14. }
\label{mpol-oxi-wc}
\end{figure}

\begin{figure}
\centerline{\psfig{file=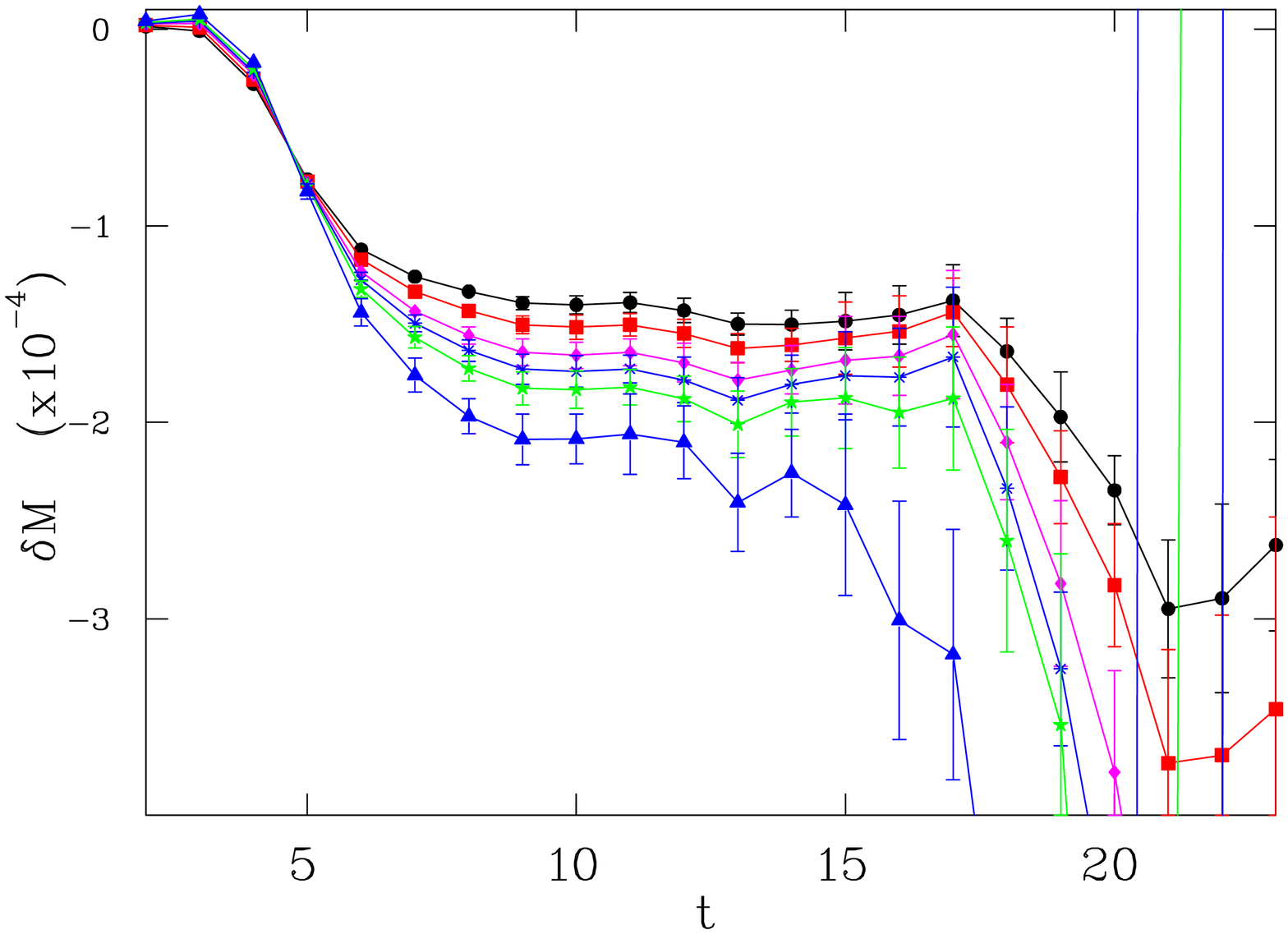,width=9cm,angle=90}}
\centerline{\psfig{file=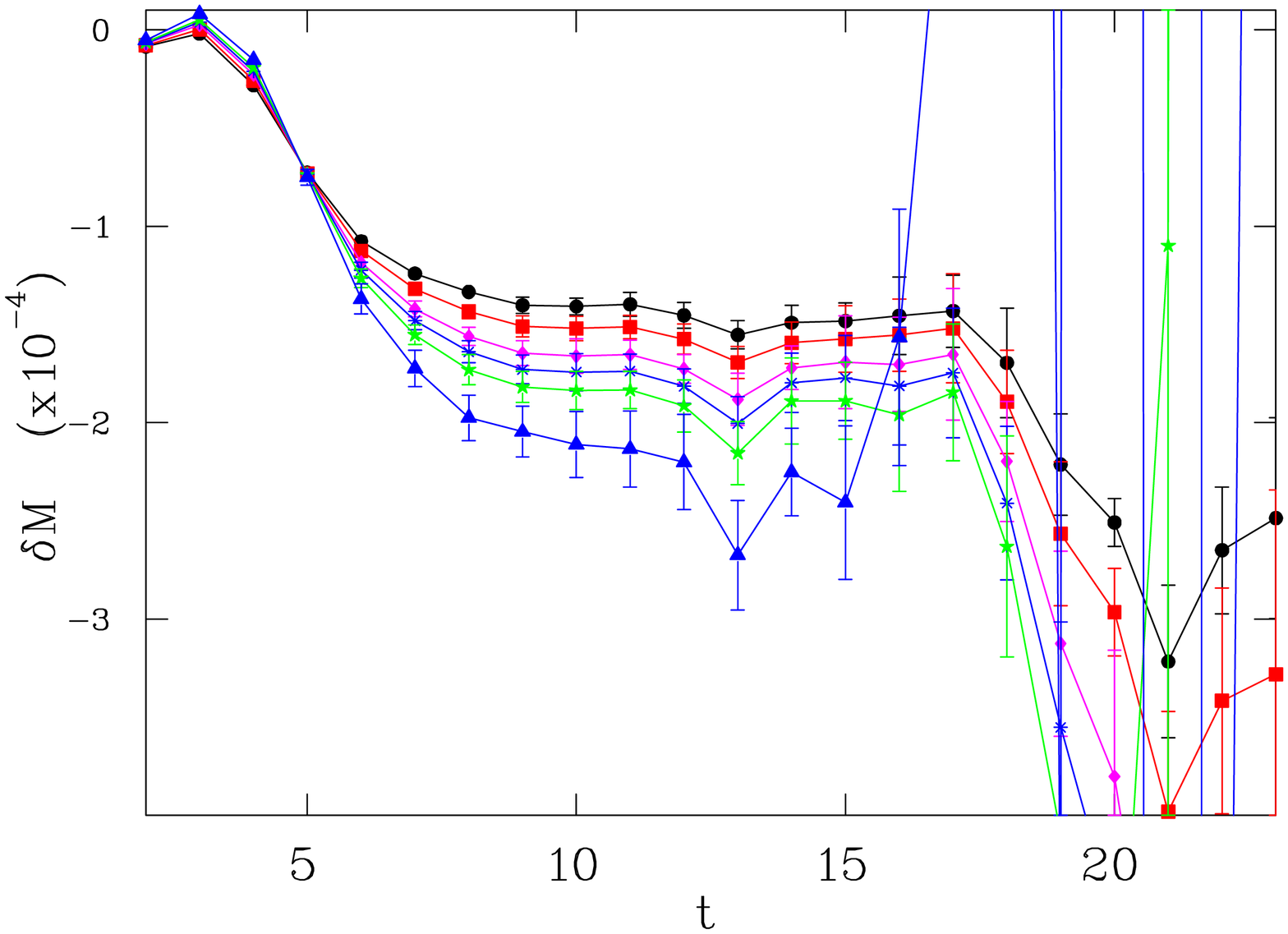,width=9cm,angle=90}}
\centerline{\psfig{file=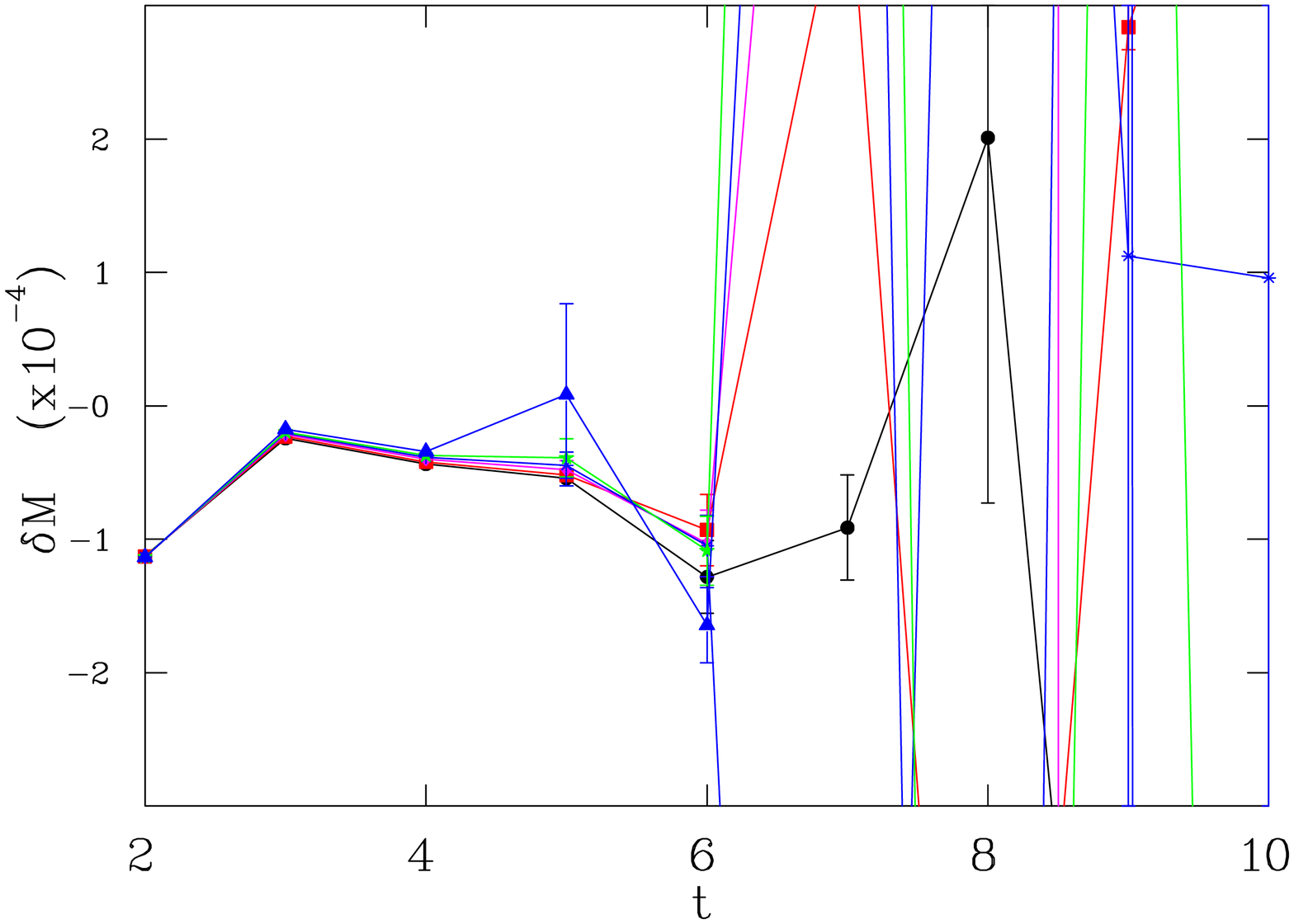,width=9cm,angle=90}}
\caption{Effective mass plots for the lambda mass shifts in lattice units
 at the weakest magnetic field 
in the order of $\Lambda^8$ (top), $\Lambda^C$ (middle), $\Lambda^S$ (bottom).
The lines correspond to quark masses from the heaviest (circles) to the lightest (triangles).}
\label{emass-olam-m1}
\end{figure}

\begin{figure}
\centerline{\psfig{file=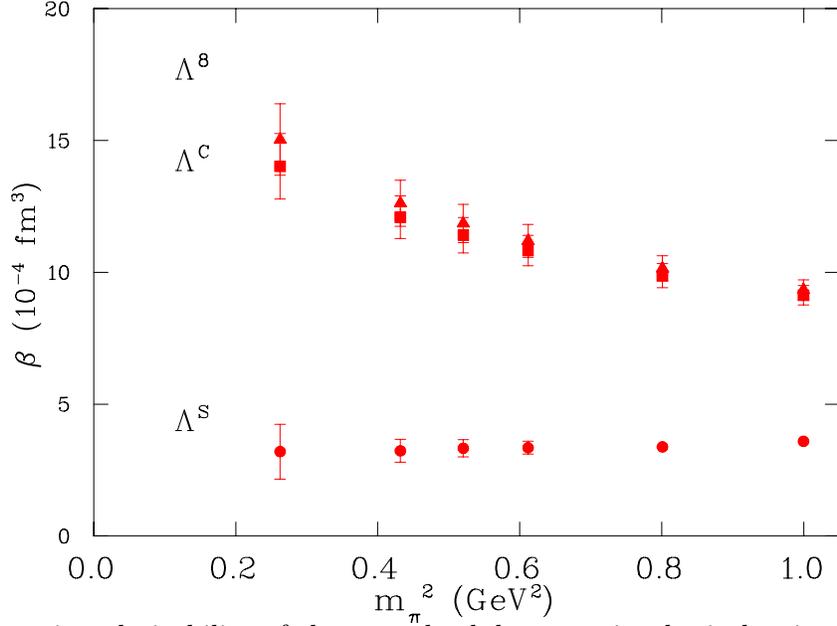,width=11cm,angle=90}}
\caption{Magnetic polarizability of the octet lambda states in physical units
from the Wilson action.
The results for $\Lambda^8$ and $\Lambda^C$ are obtained from the time window of 12 to 14, 
while for $\Lambda^S$ 5 to 7. }
\label{mpol-olam-wc}
\end{figure}

Fig.~\ref{emass-oxi-m1} shows the effective mass plot for the octet cascade states, and
Fig.~\ref{mpol-oxi-wc} shows the corresponding magnetic polarizability extracted from 12 to 14.
Fig.~\ref{emass-olam-m1} shows the effective mass plot for the octet lambda states, and
Fig.~\ref{mpol-olam-wc} shows the corresponding magnetic polarizability.
In this case, only the Wilson results are available. One can see that the results for $\Lambda^8$ and $\Lambda^C$ 
are almost the same. The signal for flavor-singlet $\Lambda^S$ plateaus much earlier than the other octet states, and the signal is lost quickly. 
Our results are extracted from the range 5 to 7.
 
The results of our calculation in the octet sector from the Wilson quark action 
are summarized in Table~\ref{mag-tab-oct}.

\begin{center}
\begin{table*}  
\caption{The calculated magnetic polarizabilities for the octet baryons as a 
function of the pion mass from the Wilson action.
The pion mass is in GeV and the magnetic polarizability is
in $10^{-4}$ fm$^3$. The time window from which each polarizability is extracted 
is given in the last column. The errors are statistical.}
 \vspace*{+0.5cm}
\label{mag-tab-oct}
\begin{tabular}{llllllll}
$\kappa$ & 0.1515 & 0.1525 & 0.1535 & 0.1540 & 0.1545 & 0.1555 &  fit range      \\
$ m_\pi$ & 1.000 & 0.895 & 0.782 & 0.721 & 0.657 & 0.512 &     \\
\hline
p            &
  0.09  $\pm$    0.29&
  0.14 $\pm$     0.37&
  0.26 $\pm$     0.48&
  0.40 $\pm$     0.56&
  0.64 $\pm$     0.67&
  2.36 $\pm$     1.20& 12-14       \\
n            &
   9.4  $\pm$     0.2 &
   10.4 $\pm$     0.3 &
   11.6 $\pm$     0.4 &
   12.3 $\pm$     0.5 &
   13.4 $\pm$     0.6 &
   17.0 $\pm$     1.1 & 12-14 \\
$\Sigma^+$   &
 -0.15  $\pm$     0.36&
  0.09  $\pm$     0.42&
  0.24  $\pm$     0.50&
  0.40  $\pm$     0.56&
  0.61  $\pm$     0.64&
  1.60  $\pm$     1.00 & 12-14           \\
$\Sigma^0$   &
   8.0  $\pm$     0.3 &
   8.5  $\pm$     0.3 &
   9.1  $\pm$     0.4 &
   9.6  $\pm$     0.5 &
   10.1 $\pm$     0.6 &
   11.9 $\pm$     0.9 & 12-14           \\
$\Sigma^-$   &
  -11.8 $\pm$      0.3 &
  -12.6 $\pm$      0.3 &
  -13.6 $\pm$      0.4 &
  -14.2 $\pm$      0.4 &
  -14.7 $\pm$      0.4 &
  -16.1 $\pm$      0.5 & 12-14           \\
$\Xi^0$      &
   10.7 $\pm$      0.3 &
   11.3 $\pm$      0.4 &
   11.9 $\pm$      0.4 &
   12.3 $\pm$      0.4 &
   12.8 $\pm$      0.5 &
   13.9 $\pm$      0.7 & 12-14           \\
$\Xi^-$      &
  -12.6 $\pm$      0.3 &
  -13.1 $\pm$      0.3 &
  -13.8 $\pm$      0.4 &
  -14.1 $\pm$      0.4 &
  -14.6 $\pm$      0.4 &
  -15.6 $\pm$      0.5 & 12-14           \\
$\Lambda^8$    &
   9.1  $\pm$     0.4 &
   9.9  $\pm$     0.5 &
   10.8 $\pm$     0.6 &
   11.4 $\pm$     0.7 &
   12.1 $\pm$     0.8 &
   14.0 $\pm$     1.2 & 12-14           \\
$\Lambda^C$    &
   9.3  $\pm$     0.4 &
   10.2 $\pm$     0.5 &
   11.2 $\pm$     0.6 &
   11.9 $\pm$     0.7 &
   12.6 $\pm$     0.9 &
   15.0 $\pm$     1.4 & 12-14           \\
$\Lambda^S$    &
   3.6  $\pm$     0.1 &
   3.4  $\pm$     0.2 &
   3.3  $\pm$     0.2 &
   3.3  $\pm$     0.3 &
   3.2  $\pm$     0.4 &
   3.2  $\pm$     1.0 & 5-7            \\
\end{tabular}
\end{table*}
\end{center}

Judging from the values at the smallest pion mass, one can observe the following 
broad features in the octet sector.
The positively-charged states ($p$ and $\Sigma^-$) have small and positive values 
around $2$, but suffer from large errors. 
The charge-neutral states ($n$, $\Sigma^0$, $\Xi^0$, 
$\Lambda^8$) have similar values to each other around $12$. 
The negatively-charged states ($\Sigma^-$, $\Xi^-$)
have similar values to each other around $-16$. 
The $\Lambda^S$ has a very weak quark mass dependence. 
The values are expected to change a little bit after chiral extrapolations.

In the case of hyperons, quark model~\cite{Lipkin92} gives $1.7$ for $\Sigma^+$ and 
$-1.7$ for $\Sigma^-$, compared with our values of about $2$ and $-16$, respectively. 
ChPT at the $O(p^3)$ order~\cite{BKM92} predicts $\alpha=10\beta$ and small and 
positive values across the octet, which is a different pattern from the results 
on the lattice. It would be interesting to see if the pattern receives significant corrections 
at the next order of ChPT.

\subsection{Decuplet Baryons}

\begin{figure}
\centerline{\psfig{file=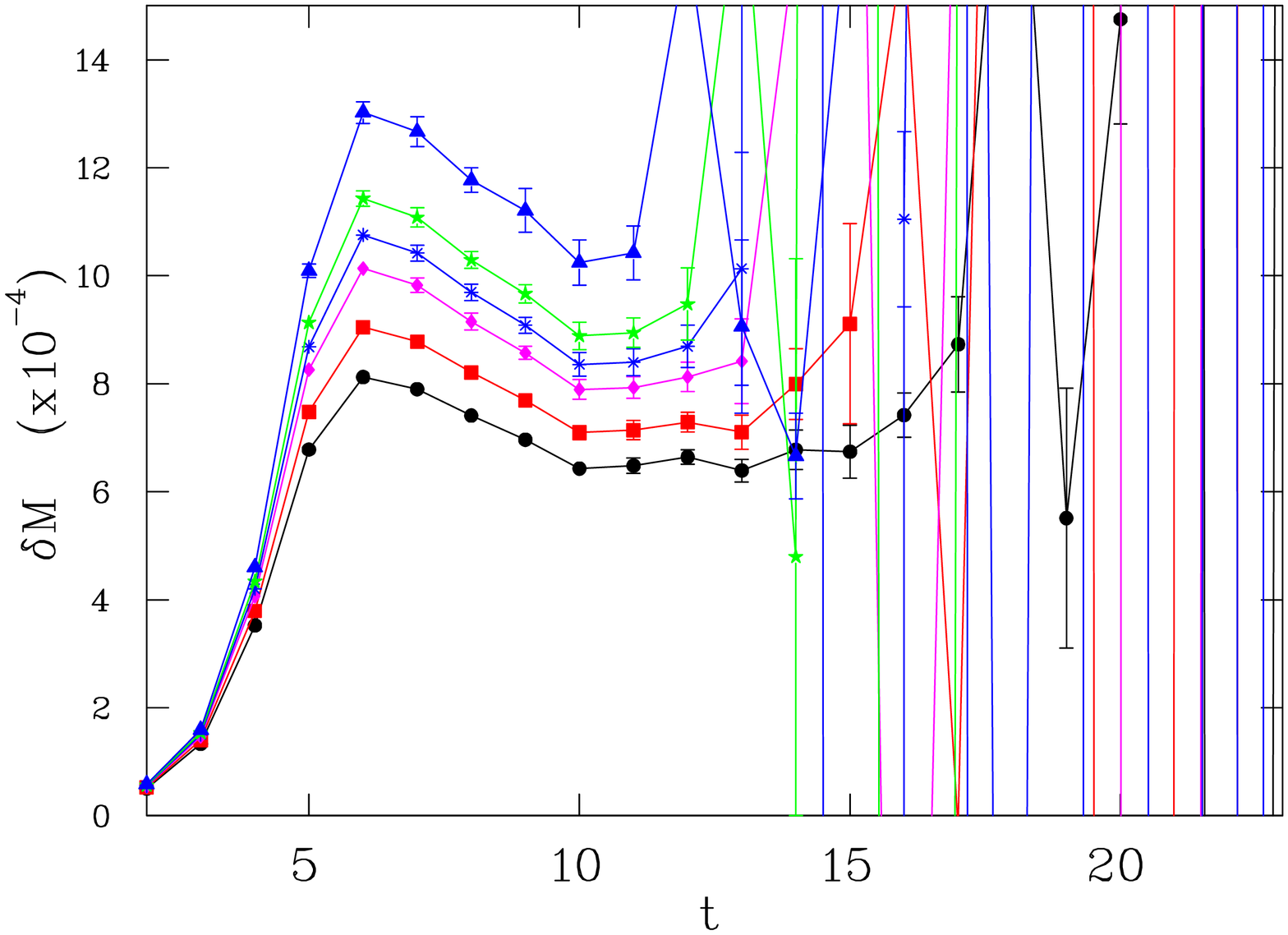,width=7.5cm,angle=90}}
\centerline{\psfig{file=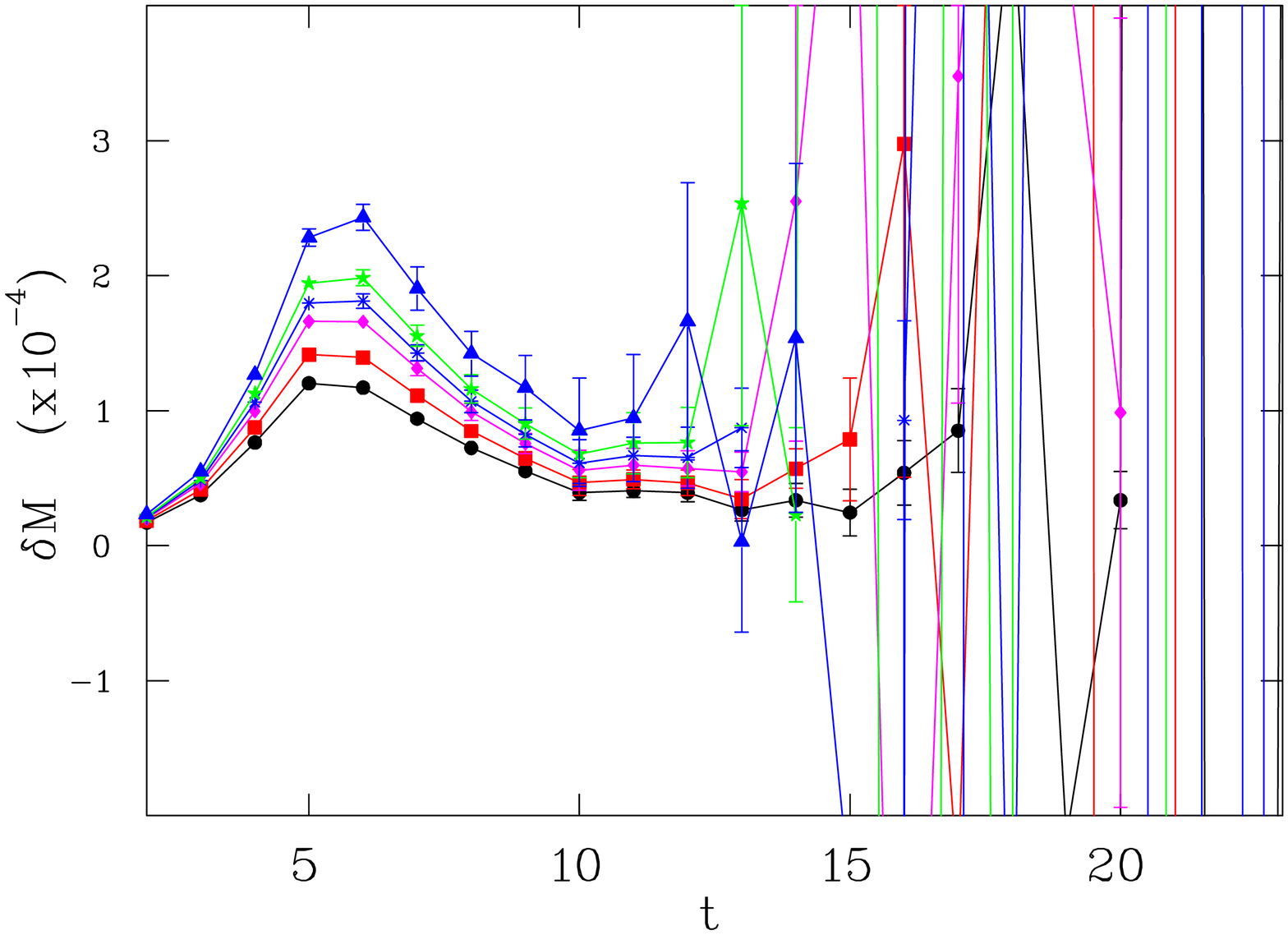,width=7.5cm,angle=90}}
\centerline{\psfig{file=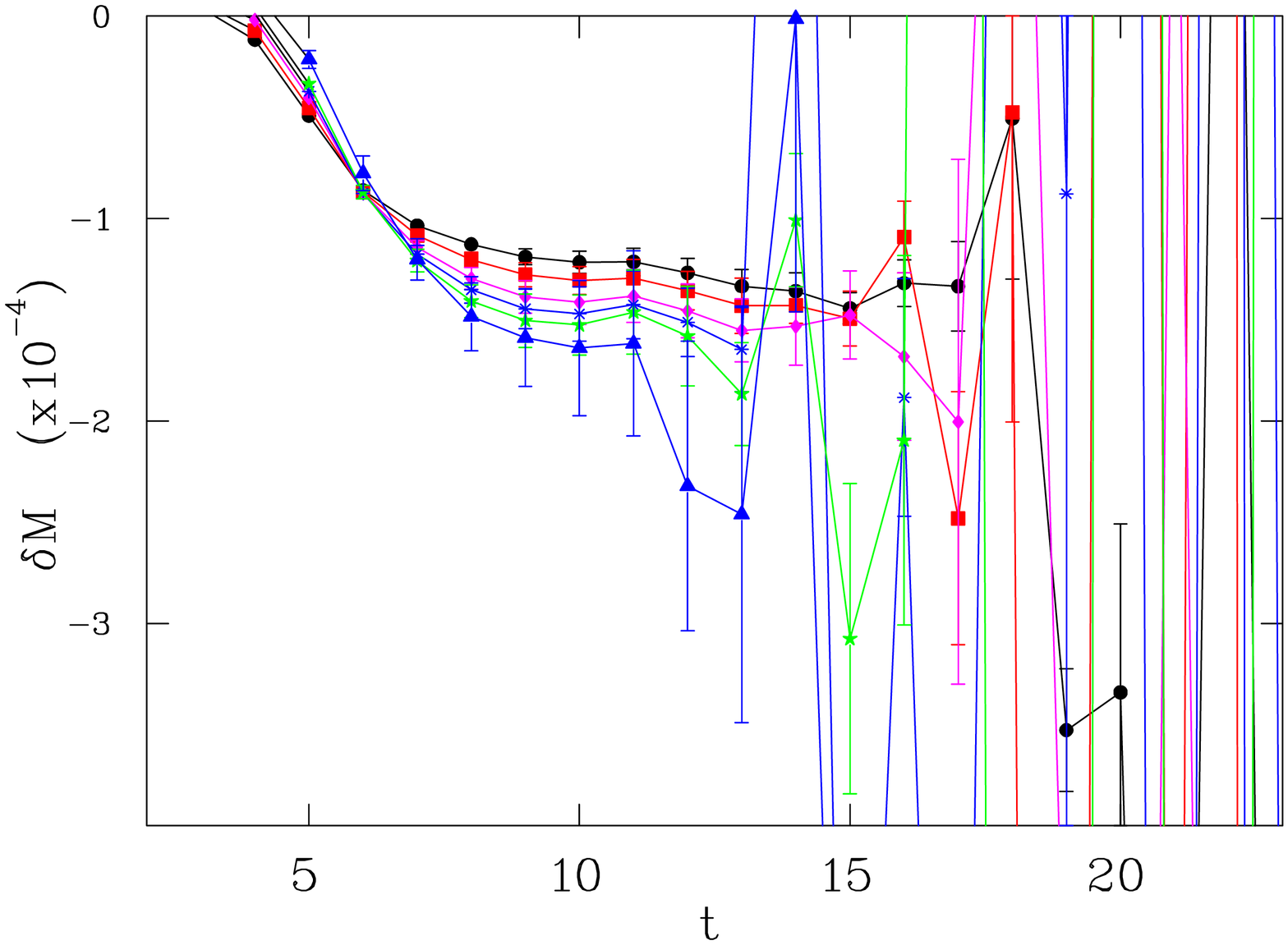,width=7.5cm,angle=90}}
\centerline{\psfig{file=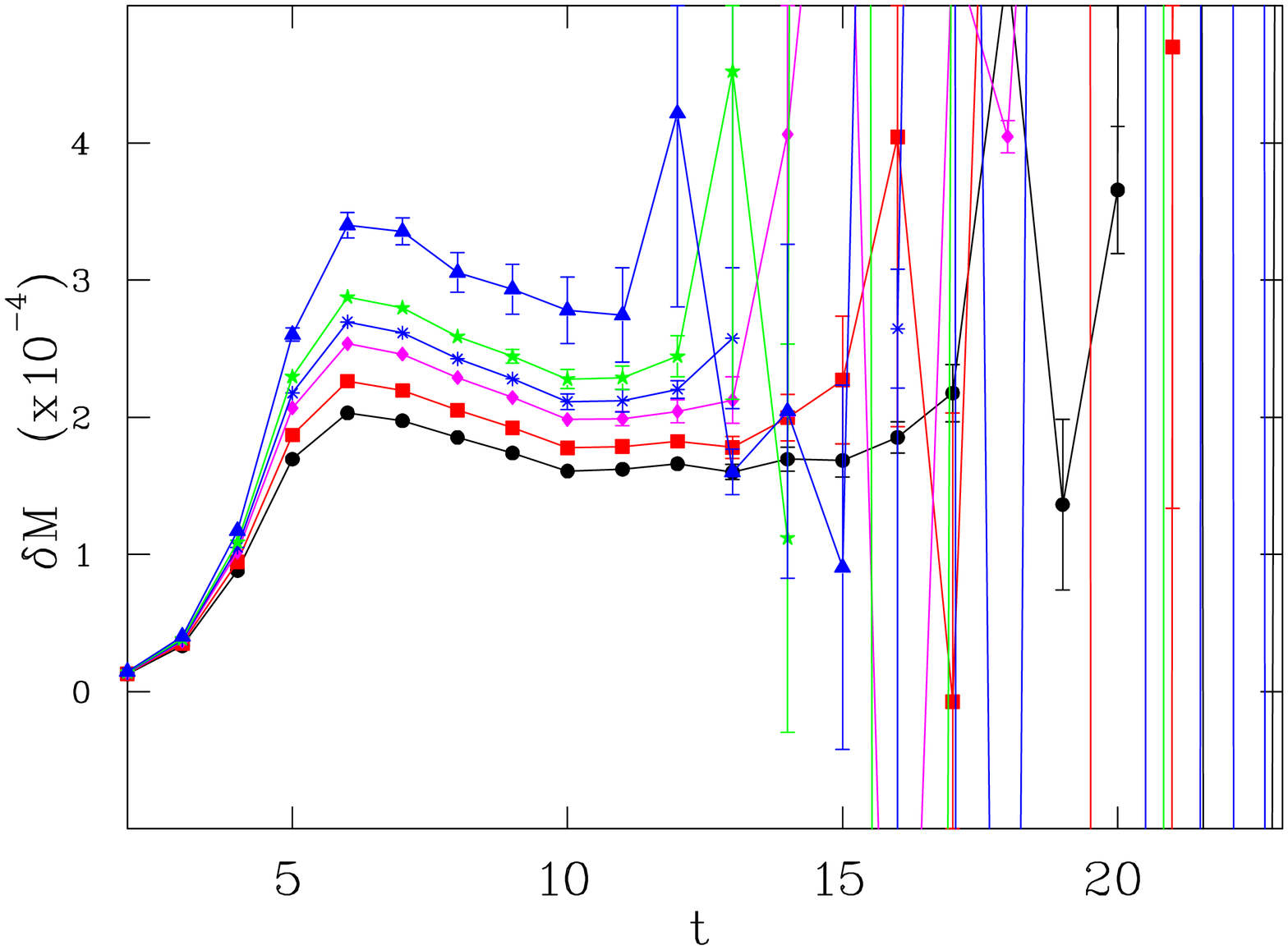,width=7.5cm,angle=90}}
\caption{Effective mass plots for the Delta states mass shifts in lattice units at the weakest magnetic field 
in the order of, from top down, $\Delta^{++}$, $\Delta^+$, $\Delta^0$, $\Delta^{-}$.
The lines correspond to quark masses from the heaviest (circles) to the lightest (triangles).}
\label{emass-decd-m1}
\end{figure}

\begin{figure}
\centerline{\psfig{file=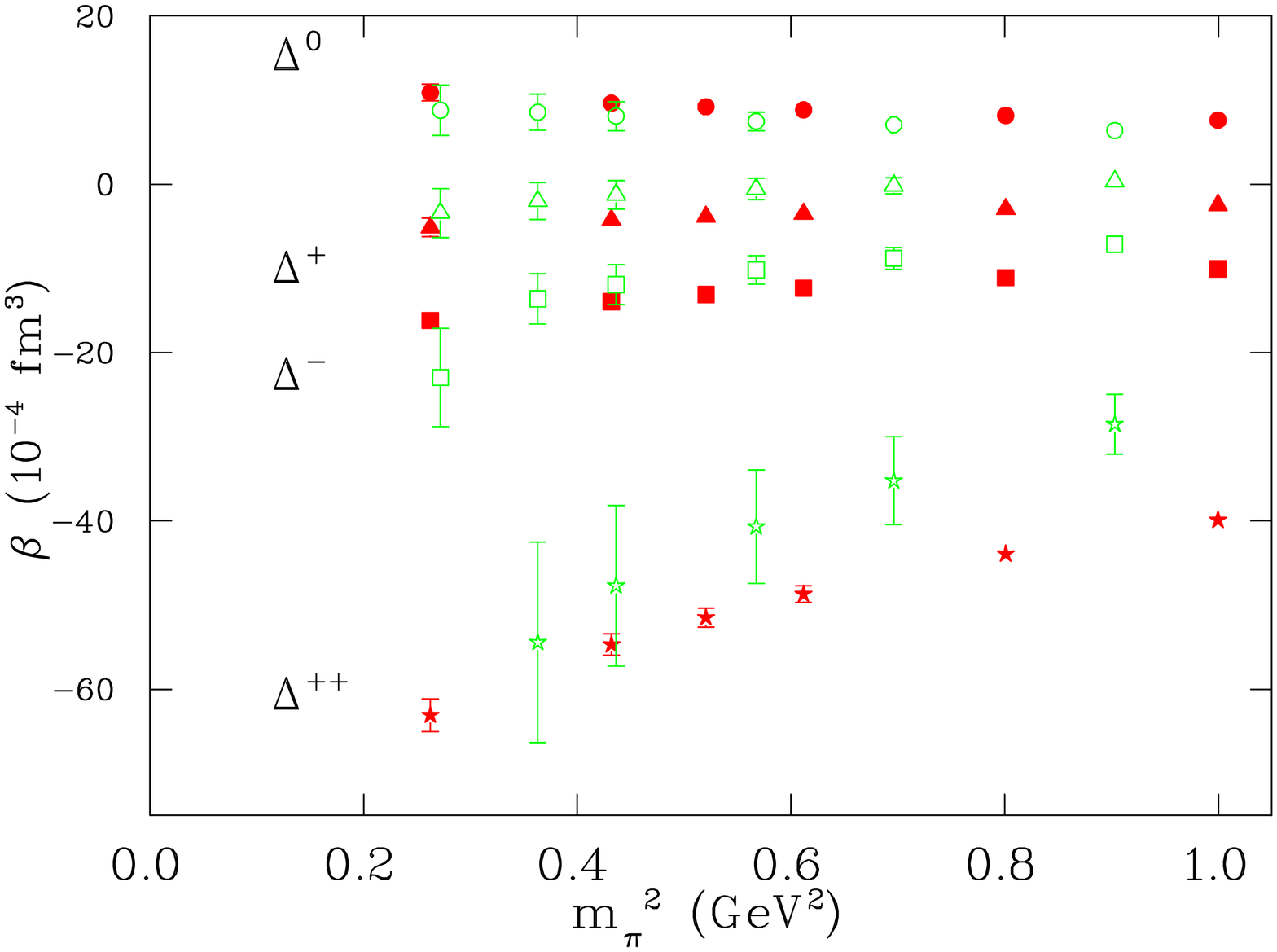,width=11cm,angle=90}}
\caption{Magnetic polarizability of the $\Delta$'s in physical units.
The solid symbols are the results from the Wilson
action, while the empty symbols are from the clover action.
The Wilson results are obtained from the time window of 10 to 12.}
\label{mpol-decd-wc}
\end{figure}

\begin{figure}
\centerline{\psfig{file=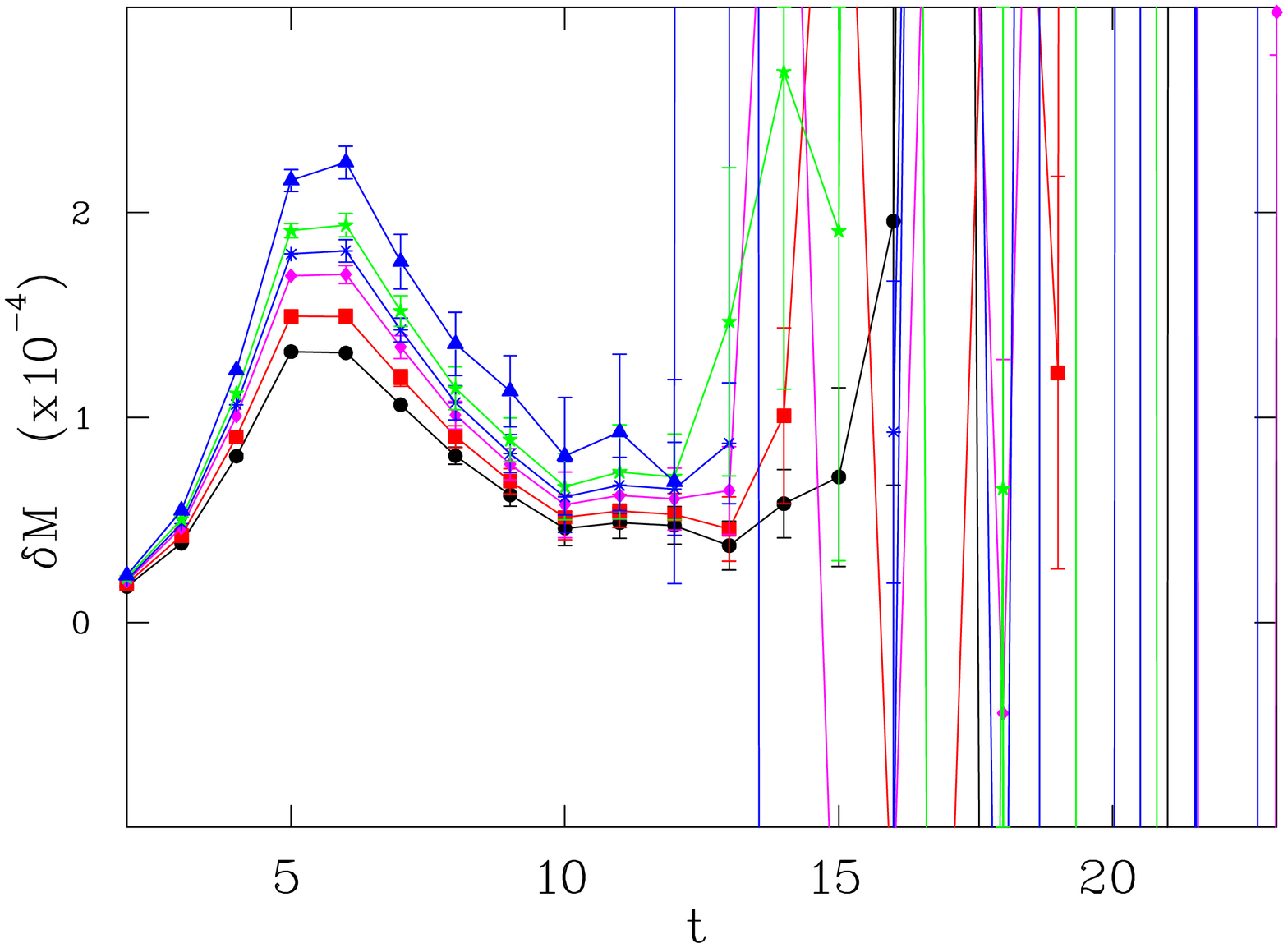,width=9cm,angle=90}}
\centerline{\psfig{file=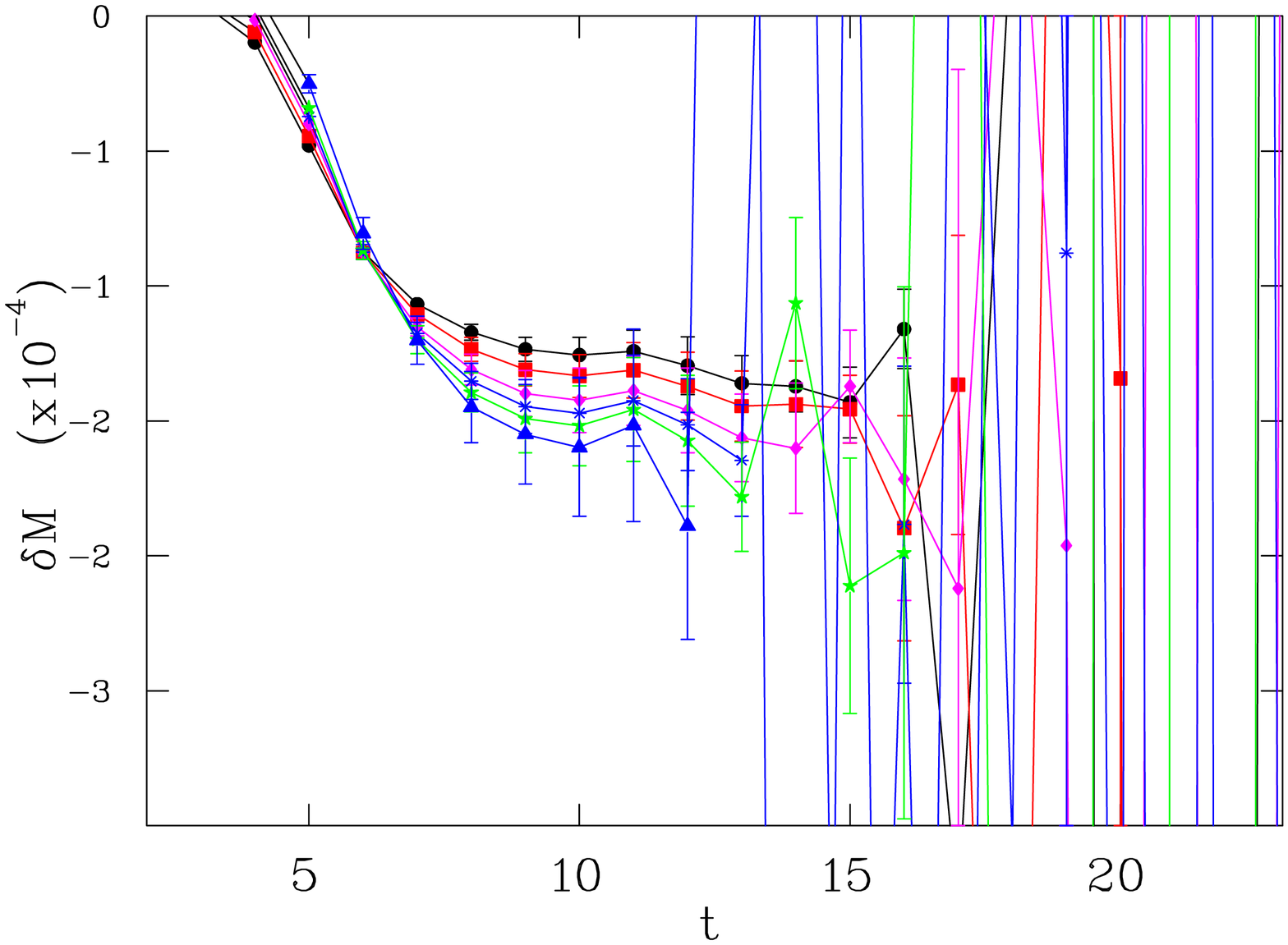,width=9cm,angle=90}}
\centerline{\psfig{file=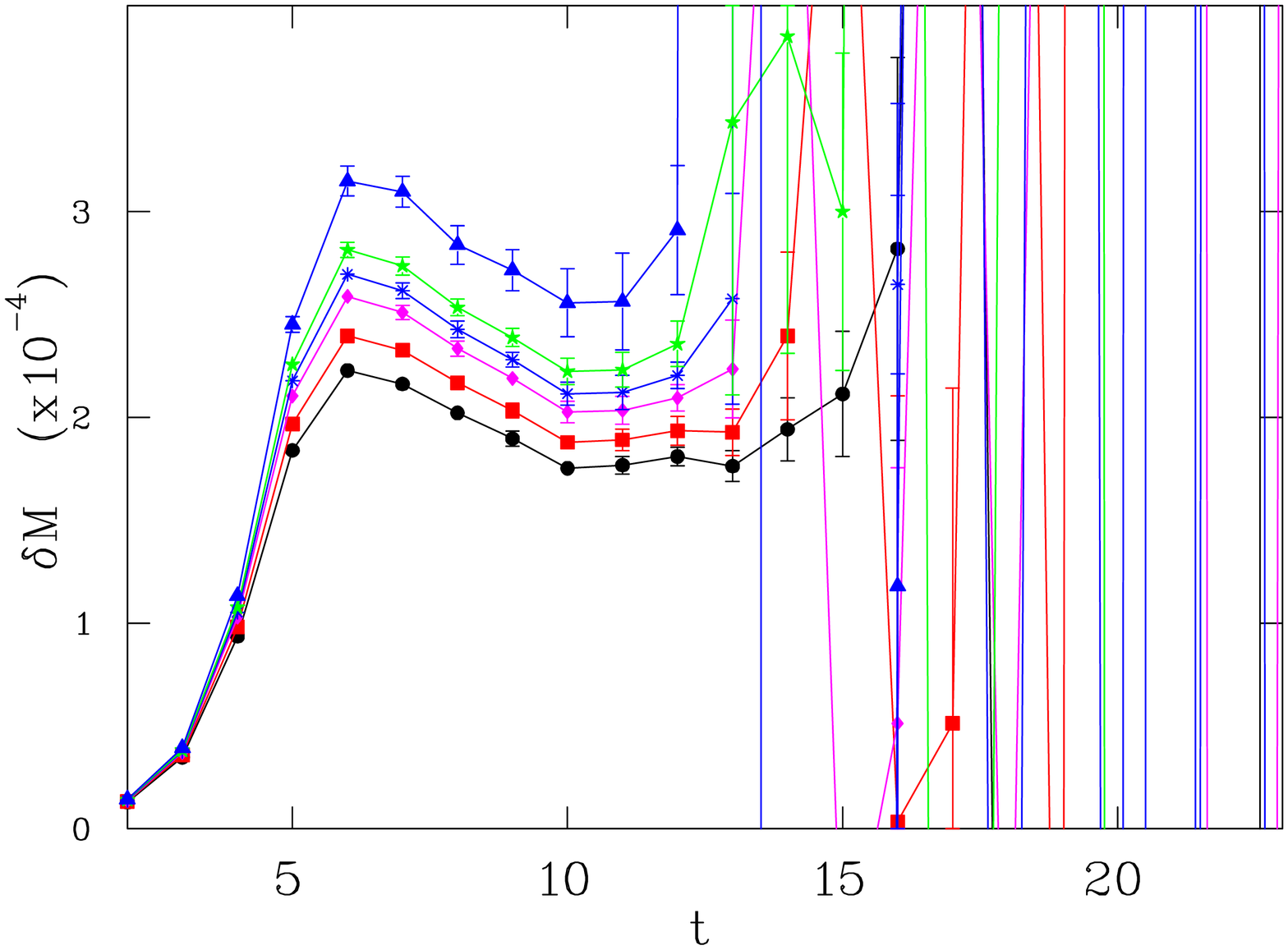,width=9cm,angle=90}}
\caption{Effective mass plots for the decuplet sigma mass shifts in lattice units 
at the weakest magnetic field in the order of 
$\Sigma^{*+}$ (top), $\Sigma^{*0}$ (middle), $\Sigma^{*-}$ (bottom).
The lines correspond to quark masses from the heaviest (circles) to the lightest (triangles).}
\label{emass-dsig-m1}
\end{figure}

\begin{figure}
\centerline{\psfig{file=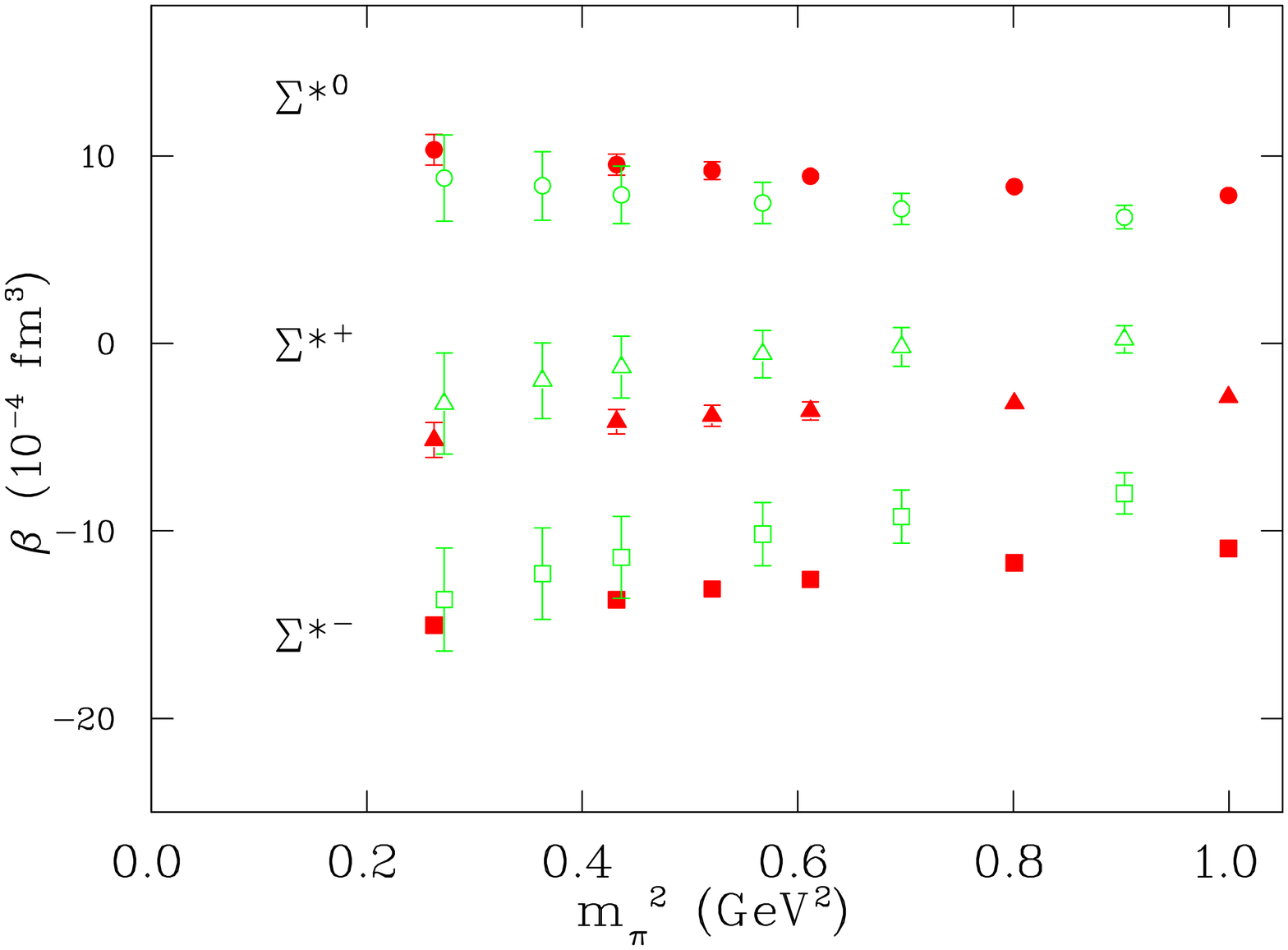,width=11cm,angle=90}}
\caption{Magnetic polarizability of the decuplet sigma states in physical units.
The solid symbols are the results from the Wilson
action, while the empty symbols are from the clover action.
The Wilson results are obtained from the time window of 10 to 12. }
\label{mpol-dsig-wc}
\end{figure}

Fig.~\ref{emass-decd-m1} displays the effective mass plot for the Delta states.
The plateaus form earlier for the decuplet states than the octet states, 
probably due to 
the fact that the decuplet states are heavier.
Fig.~\ref{mpol-decd-wc} shows the magnetic polarizability for the four Delta states. 
The interesting result here is that the $\Delta^{++}$ polarizability is large and 
negative (around -60 at the smallest pion mass), 
which means it is most easily deformed (decreasing mass) 
under the probing magnetic field.
It is the largest magnetic polarizability we observed on the lattice.

\begin{figure}
\centerline{\psfig{file=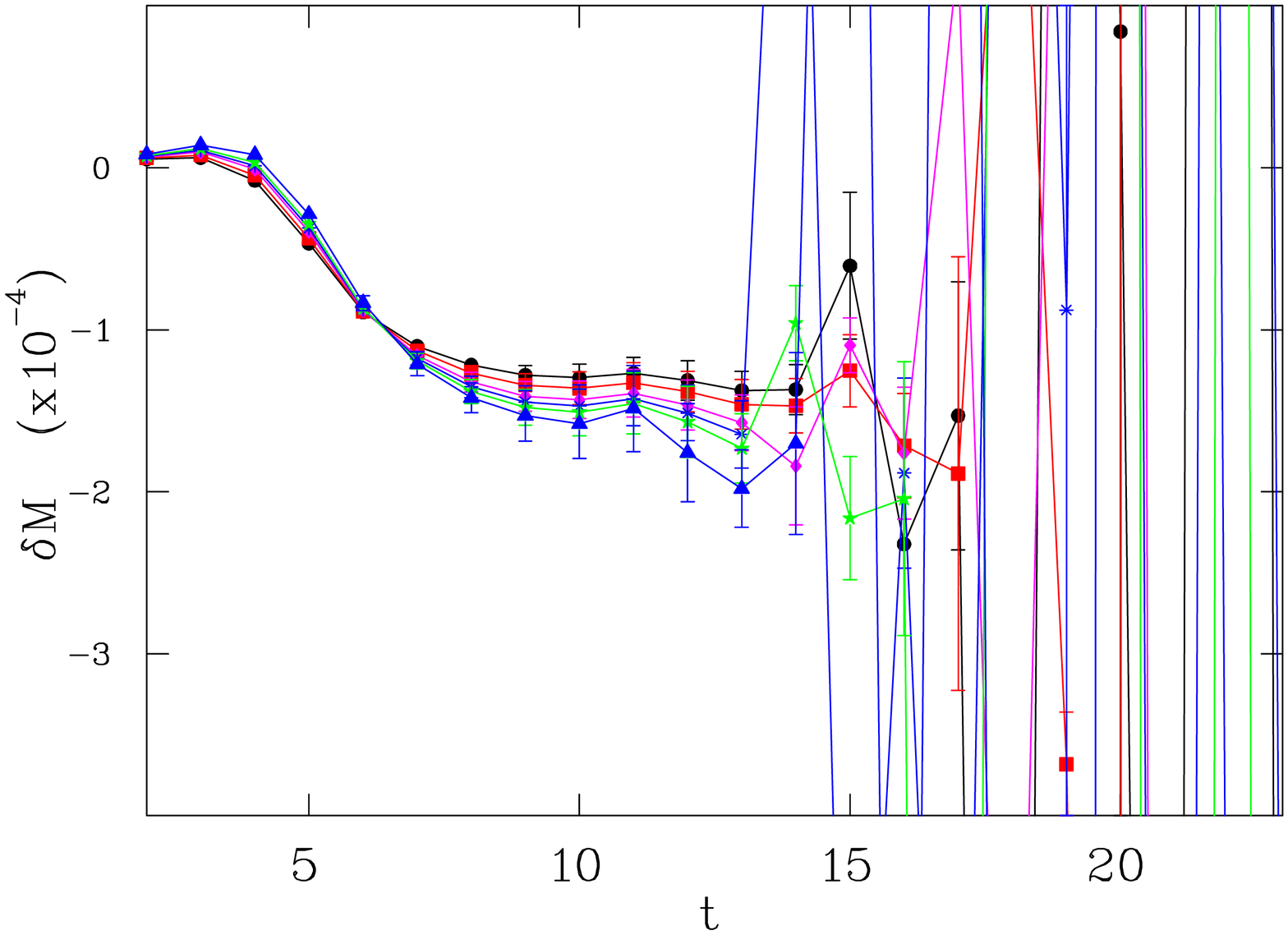,width=7.5cm,angle=90}}
\centerline{\psfig{file=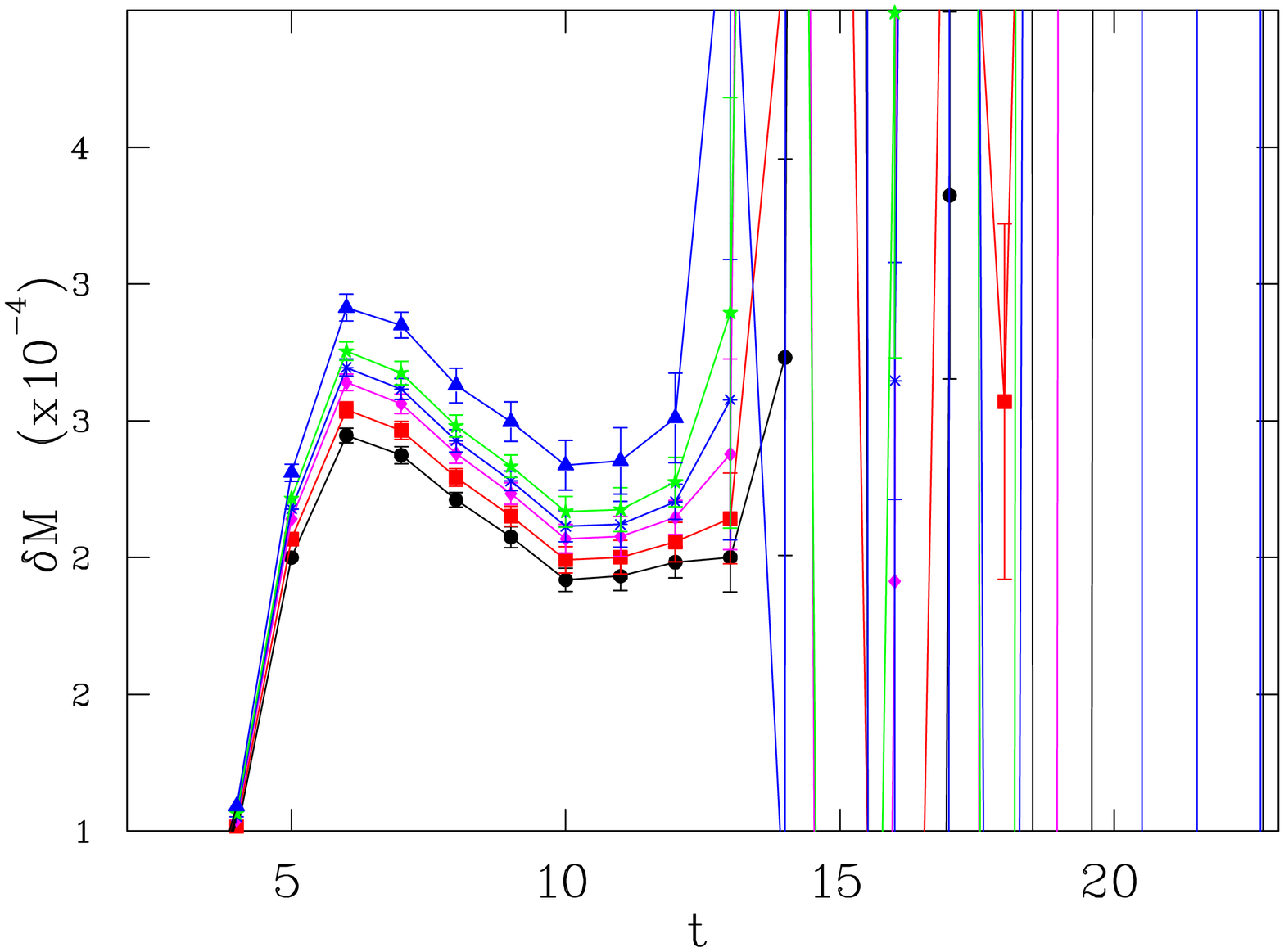,width=7.5cm,angle=90}}
\caption{Effective mass plots for the decuplet $\Xi^{*0}$ (upper), $\Xi^{*-}$ (lower) 
mass shifts in lattice units at the weakest magnetic field.
The lines correspond to quark masses from the heaviest (circles) to the lightest (triangles).}
\label{emass-dxi-m1}
\end{figure}

\begin{figure}
\centerline{\psfig{file=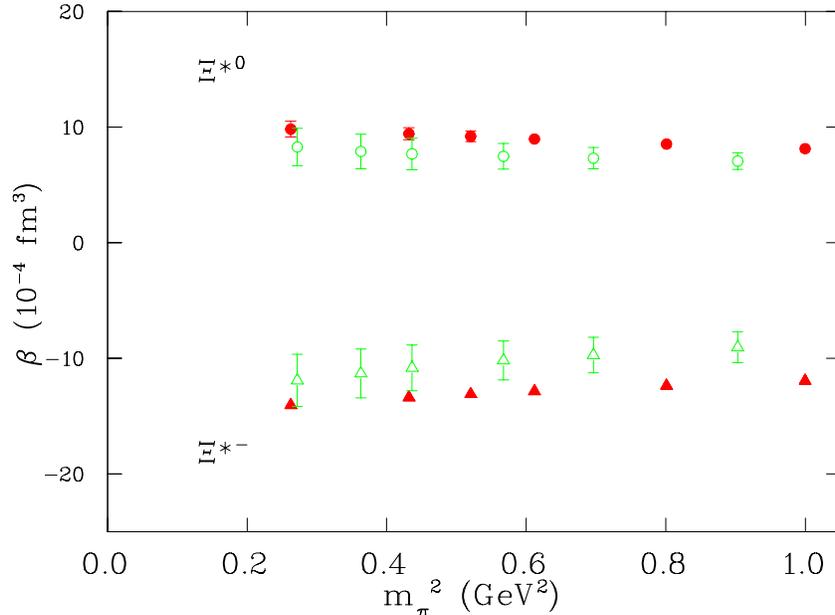,width=11cm,angle=90}}
\caption{Magnetic polarizability of the decuplet cascade states in physical units.
The solid symbols are the results from the Wilson
action, while the empty symbols are from the clover action.
The Wilson results are obtained from the time window of 10 to 12.}
\label{mpol-dxi-wc}
\end{figure}

Fig.~\ref{emass-dsig-m1} displays the effective mass plot for the decuplet sigma states, 
and Fig.~\ref{mpol-dsig-wc} shows the corresponding magnetic polarizability as a function 
of the pion mass squared.
Fig.~\ref{emass-dxi-m1} displays the effective mass plot for the decuplet cascade states, 
and Fig.~\ref{mpol-dxi-wc} shows the corresponding magnetic polarizability as a function 
of the pion mass squared.

The results of our calculation in the decuplet sector from the Wilson action are
summarized in Table~\ref{mag-tab-dec}.

From the values at the smallest pion mass, one can observe the following
features in the decuplet sector.
The values are expected to change a little bit after chiral extrapolations.
The positively-charged states ($\Delta^+$ and $\Sigma^{*+}$) have negative 
values around -5 and relatively small errors. The situation is opposite to that for 
the octet members ($p$ and $\Sigma^+$).
The charge-neutral states ($\Delta^0$, $\Sigma^{*0}$, $\Xi^{*0}$)
have similar values to each other around 10, similar to the situation in the octet sector.
The negatively-charged states ($\Sigma^{*-}$, $\Xi^{*-}$, $\Omega^{-}$)
have similar values to each other around -15, again similar to the situation in the octet sector.
Note that the value for $\Omega^{-}$, -12.4(2), is simply predicted from the $\Delta^-$ 
at the strange quark mass ($\kappa$=0.1535). This prediction is free of uncertainty  
from chiral extrapolations. An experimental measurement of the $\Omega^{-}$ 
magnetic polarizability is greatly desired. Historically, the precise determination of 
the mass and magnetic moment of the $\Omega^{-}$ has paid significant dividends to our 
understanding of hadron structure and dynamics. Its magnetic polarizability provides 
another such potential opportunity.

There is no experimental measurements on the decuplet states, 
and very limited theoretical information.
An estimate in the Skyrme model~\cite{Scherer92} gives a value of -14.9 for $\Delta^-$, 
which is consistent with the value on the lattice.
The large and negative value for the $\Delta^{++}$ calls for theoretical explanations.

\begin{center}
\begin{table*}  
\caption{The calculated magnetic polarizabilities for the decuplet baryons
as a function of the pion mass from the Wilson action.
The pion mass is in GeV and the magnetic polarizability is
in $10^{-4}$ fm$^3$. The time window from which each polarizability is extracted 
is given in the last column. The errors are statistical.}
 \vspace*{+0.5cm}
\label{mag-tab-dec}
\begin{tabular}{llllllll}
$\kappa$ & 0.1515 & 0.1525 & 0.1535 & 0.1540 & 0.1545 & 0.1555 &  fit range      \\
$ m_\pi$ & 1.000 & 0.895 & 0.782 & 0.721 & 0.657 & 0.512 &     \\
\hline
$\Delta^{++}$&
  -39.9 $\pm$  0.7&
  -43.9 $\pm$  0.8&
  -48.7 $\pm$  1.0&
  -51.5 $\pm$  1.1&
  -54.7 $\pm$  1.3&
  -63.1 $\pm$  1.9& 10-12  \\
$\Delta^{+}$&
  -2.5 $\pm$ 0.2 &
  -3.0 $\pm$ 0.3 &
  -3.5 $\pm$ 0.5 &
  -3.9 $\pm$ 0.6 &
  -4.3 $\pm$ 0.7 &
  -5.1 $\pm$ 1.1 & 10-12  \\
$\Delta^{0 }$&
   7.6 $\pm$  0.2&
   8.2 $\pm$  0.3&
   8.8 $\pm$  0.4&
   9.2 $\pm$  0.5&
   9.6 $\pm$  0.6&
   10.9$\pm$  1.0& 10-12           \\
$\Delta^{- }$&
  -10.1  $\pm$     0.2 &
  -11.1  $\pm$     0.2 &
  -12.4  $\pm$     0.2 &
  -13.1  $\pm$     0.3 &
  -14.0  $\pm$     0.3 &
  -16.2  $\pm$     0.5 & 10-12          \\
$\Sigma^{*+}$  &
  -2.9   $\pm$    0.3 &
  -3.2   $\pm$    0.4&
  -3.6   $\pm$    0.5&
  -3.9   $\pm$    0.6&
  -4.2   $\pm$    0.6&
  -5.1   $\pm$    0.9& 10-12  \\
$\Sigma^{*0}$  &
   7.9 $\pm$      0.2 &
   8.4 $\pm$      0.3 &
   8.9 $\pm$      0.4 &
   9.2 $\pm$      0.5 &
   9.5 $\pm$      0.6 &
   10.3 $\pm$     0.8 & 10-12           \\
$\Sigma^{*-}$  &
  -10.9   $\pm$   0.2 &
  -11.7   $\pm$   0.2 &
  -12.6   $\pm$   0.3 &
  -13.1   $\pm$   0.3 &
  -13.7   $\pm$   0.3 &
  -15.0   $\pm$   0.4 & 10-12          \\
$\Xi^{*0}$     &
   8.1 $\pm$      0.3 &
   8.5 $\pm$      0.3 &
   9.0 $\pm$      0.4 &
   9.2 $\pm$      0.5 &
   9.4 $\pm$      0.5 &
   9.8 $\pm$      0.7 & 10-12        \\
$\Xi^{*-}$     &
  -11.9  $\pm$    0.2 &
  -12.4  $\pm$    0.2 &
  -12.8  $\pm$    0.3 &
  -13.1  $\pm$    0.3 &
  -13.4  $\pm$    0.3 &
  -14.0  $\pm$    0.3 & 10-12          \\
$\Omega^{-}$     &
&&
  -12.4  $\pm$     0.2 &
&&
                      & 10-12          \\
\end{tabular}
\end{table*}
\end{center}

\subsection{Mesons}

\begin{figure}
\centerline{\psfig{file=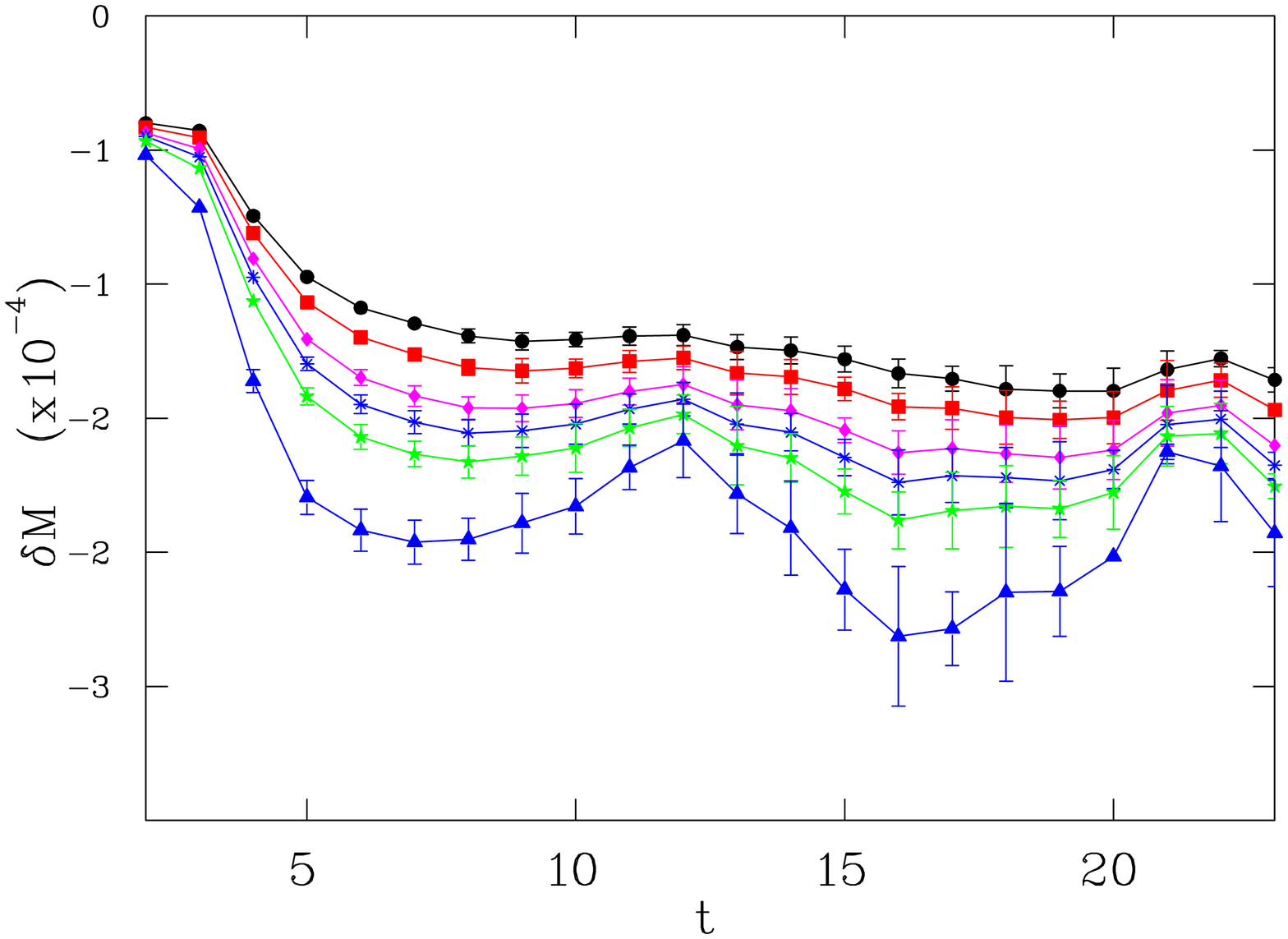,width=7.5cm,angle=90}}
\centerline{\psfig{file=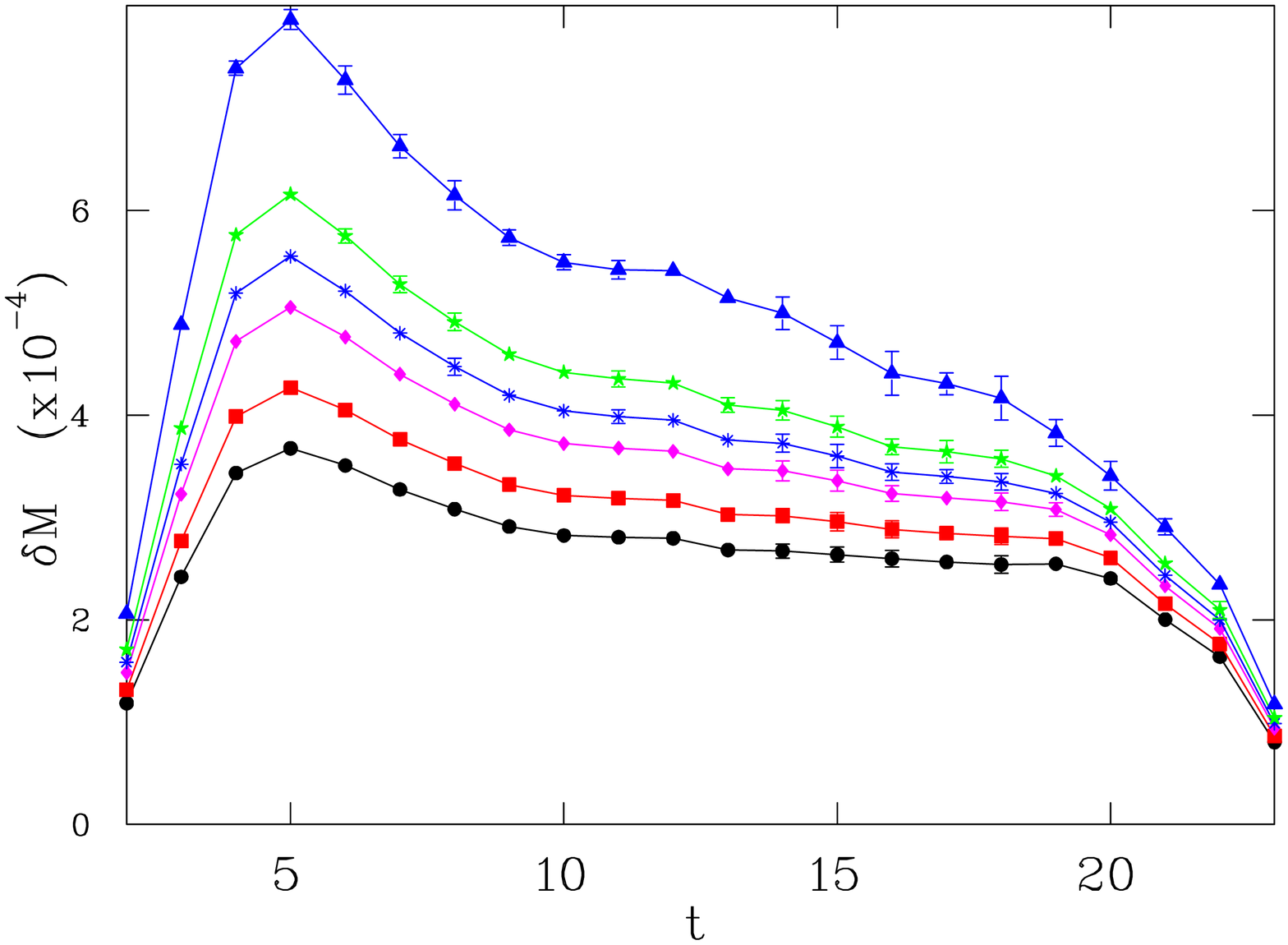,width=7.5cm,angle=90}}
\caption{Effective mass plots for the pion mass shifts in lattice units 
at the weakest magnetic field 
in the order of $\pi^0$ (upper) and $\pi^\pm$ (lower). 
The lines correspond to quark masses from the heaviest (circles) to the lightest (triangles).}
\label{emass-pi-m1}
\end{figure}

\begin{figure}
\centerline{\psfig{file=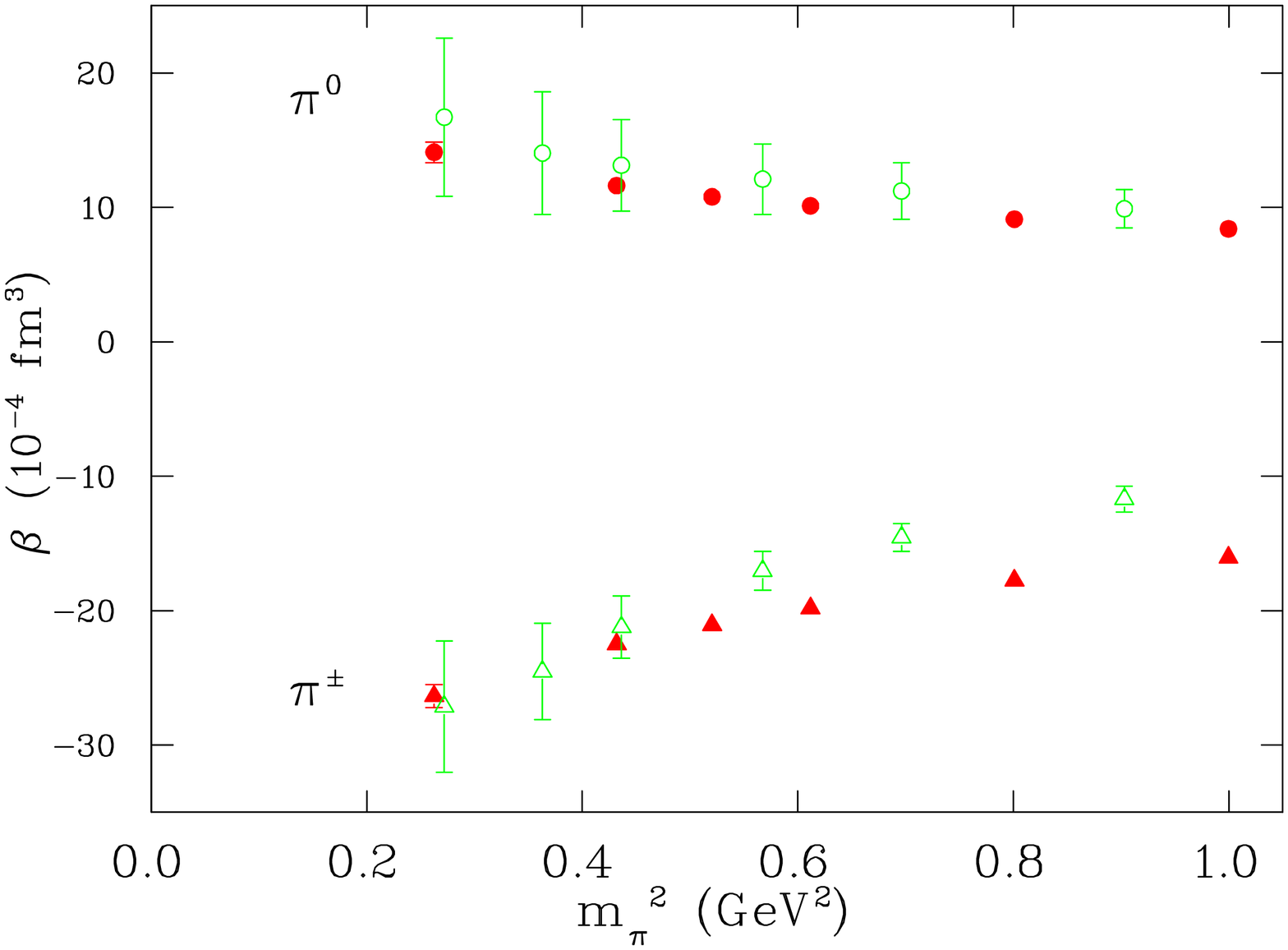,width=11cm,angle=90}}
\caption{Magnetic polarizability for the pion states in physical units.
The solid symbols are the results from the Wilson
action, while the empty symbols are from the clover action.
The Wilson results are obtained from the time window of 16 to 18. }
\label{mpol-pi-wc}
\end{figure}

Fig.~\ref{emass-pi-m1} displays the effective mass plot for the pion states.
The plateau happens much later than in the neutron.
We fit in the range of 16 to 18.
In our calculation, $\pi^+$ and $\pi^-$ have identical mass shifts, meaning they have 
identical magnetic polarizability. The same is true for all other charged mesons we considered.
This is expected because they are anti-particles to each other and have the same mass.
Fig.~\ref{mpol-pi-wc} shows the corresponding magnetic polarizability as a function 
of the pion mass squared.

\begin{figure}
\centerline{\psfig{file=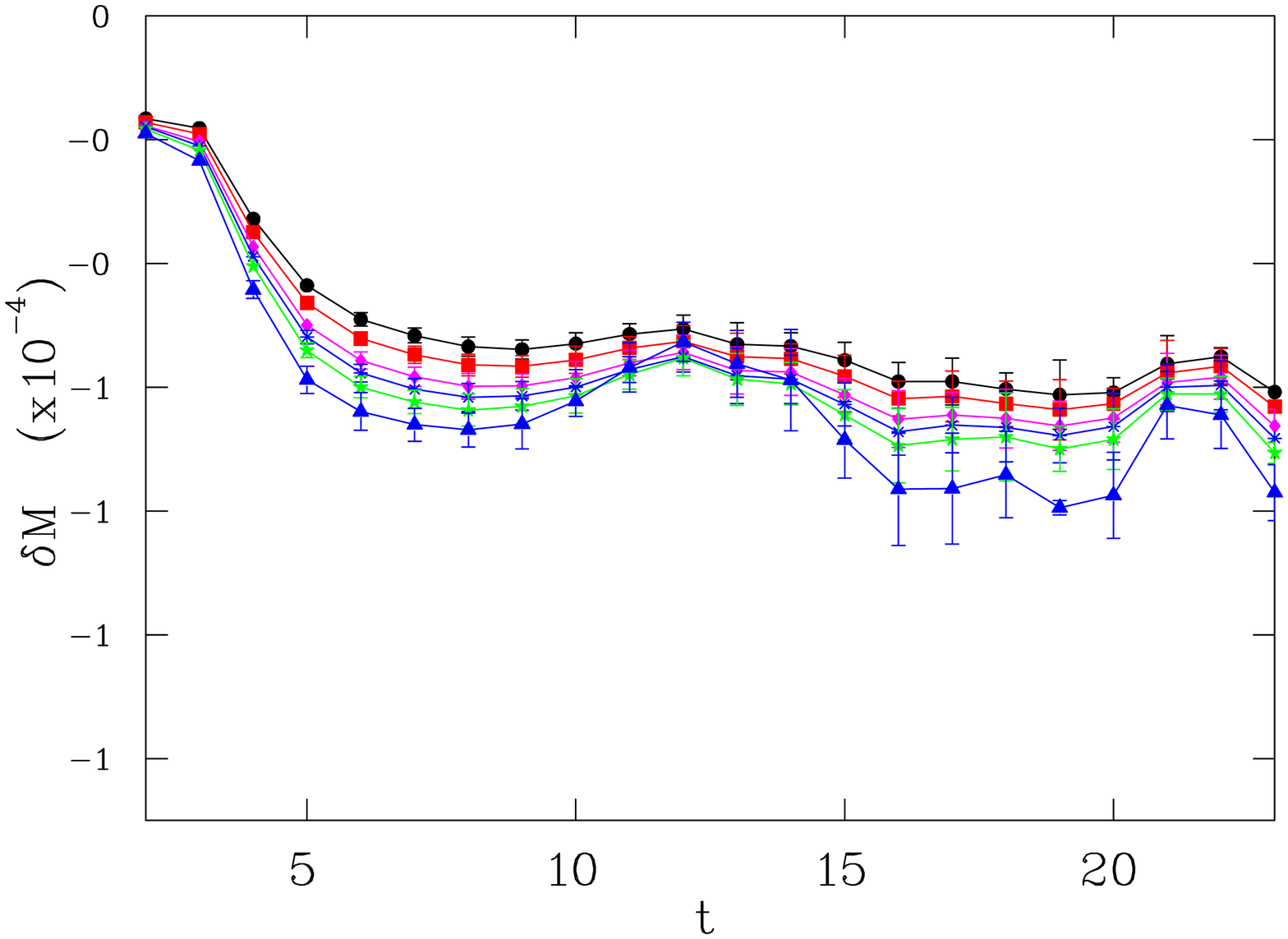,width=7.5cm,angle=90}}
\centerline{\psfig{file=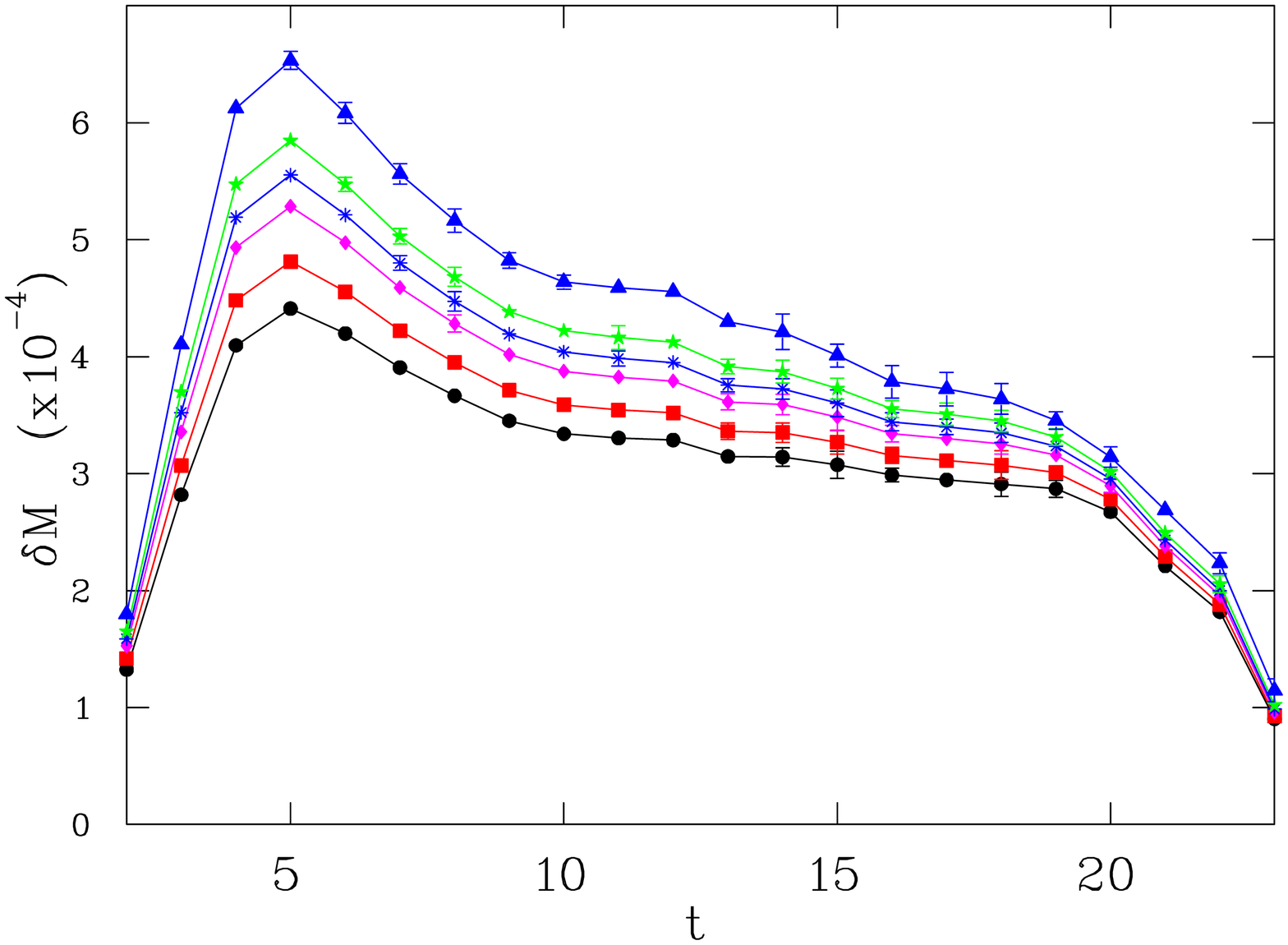,width=7.5cm,angle=90}}
\caption{Effective mass plots for the kaon mass shifts in lattice units 
at the weakest magnetic field 
in the order of $K^0$ (upper) and $K^\pm$ (lower). 
The lines correspond to quark masses from the heaviest (circles) to the lightest (triangles).}
\label{emass-K-m1}
\end{figure}

\begin{figure}
\centerline{\psfig{file=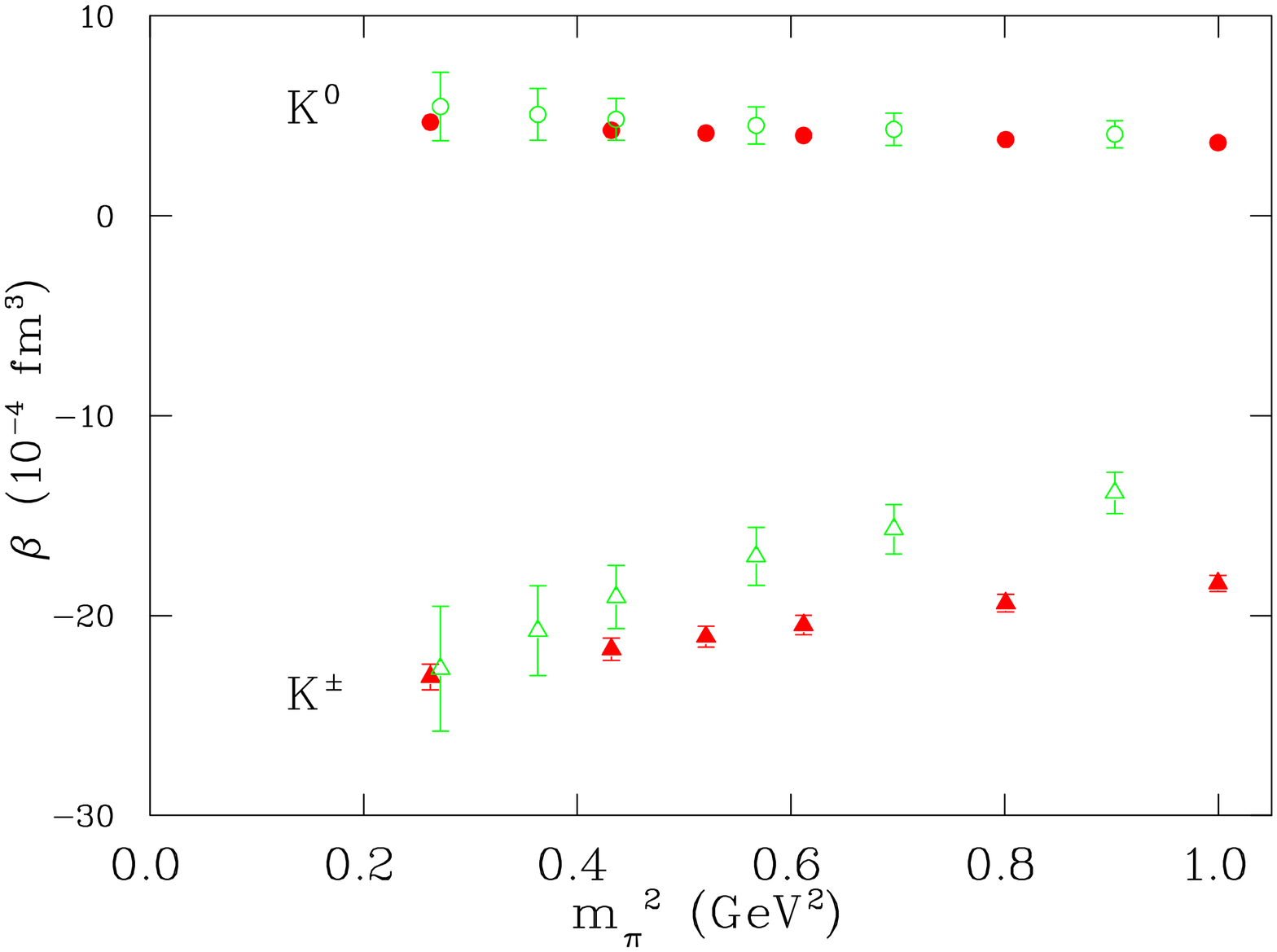,width=11cm,angle=90}}
\caption{Magnetic polarizability for the kaon states in lattice units.
The solid symbols are the results from the Wilson
action, while the empty symbols are from the clover action.
The Wilson results are obtained from the time window of 16 to 18. }
\label{mpol-K-wc}
\end{figure}

Fig.~\ref{emass-K-m1} displays the effective mass plot for the kaon states.
The quality of the plateaus is about the same as that of the pion states 
and we fit in the same range of 16 to 18.
Fig.~\ref{mpol-K-wc} the corresponding magnetic polarizability as a function 
of the pion mass squared.

\begin{figure}
\centerline{\psfig{file=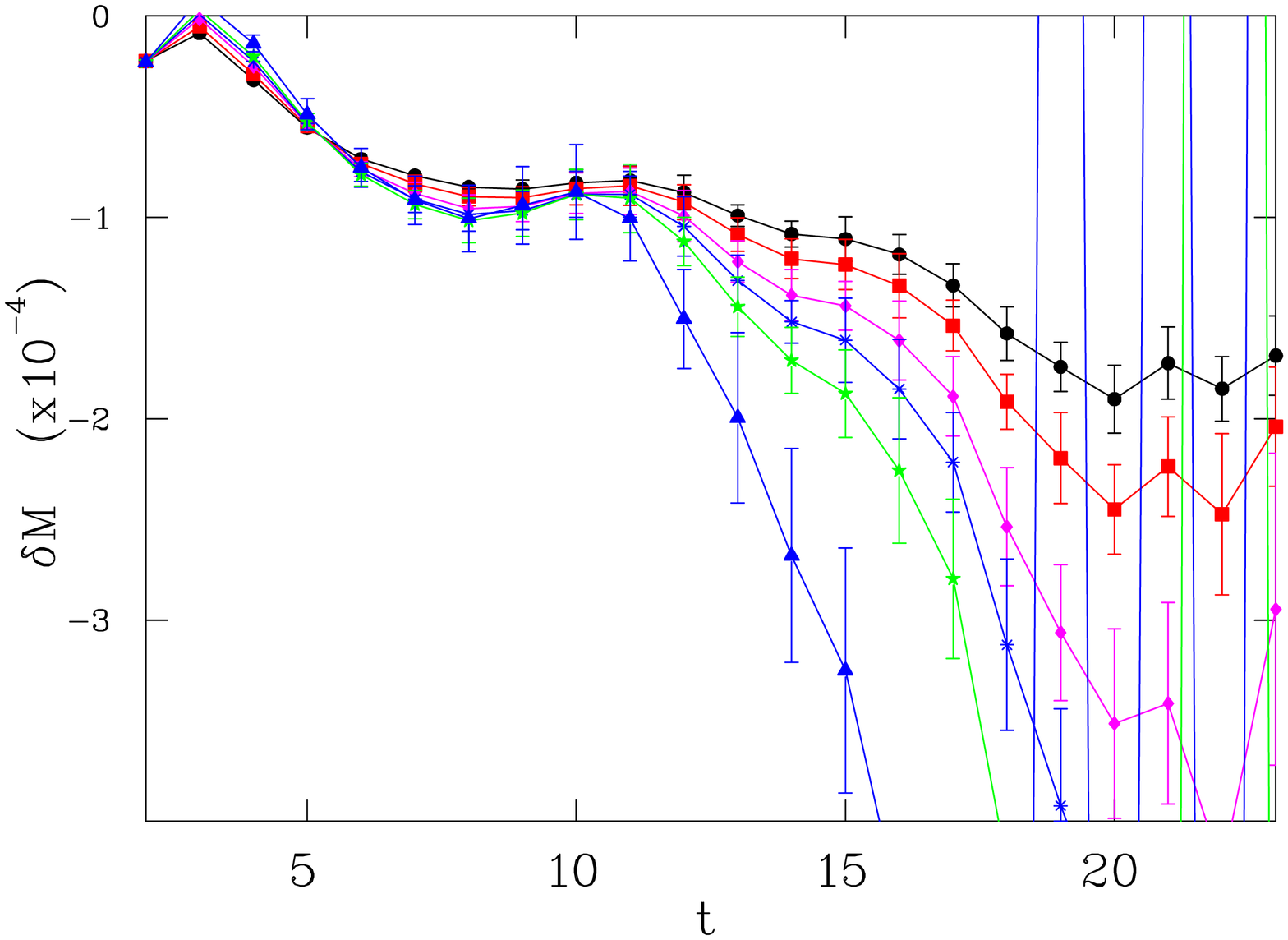,width=7.5cm,angle=90}}
\centerline{\psfig{file=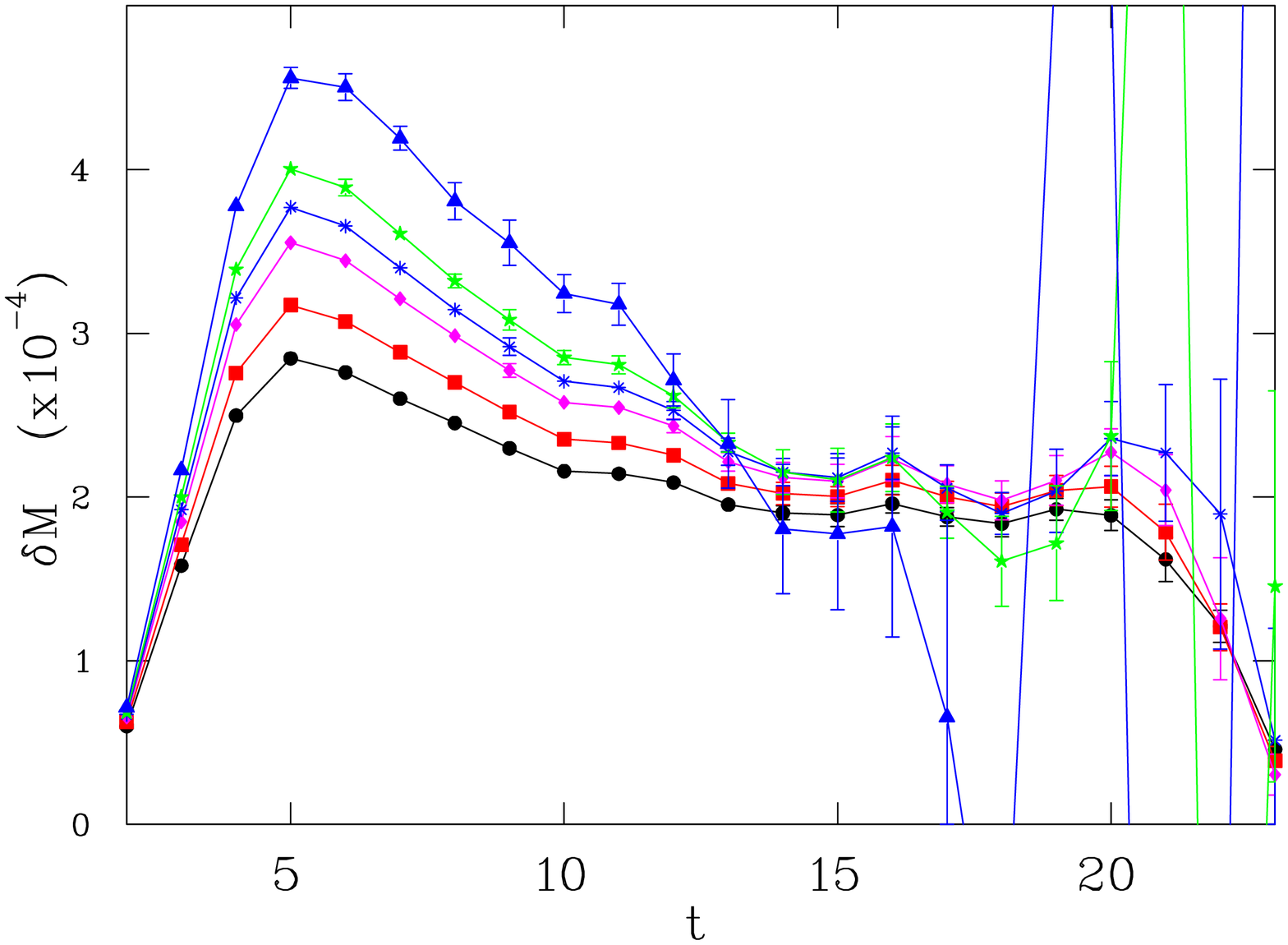,width=7.5cm,angle=90}}
\caption{Effective mass plots for the rho meson mass shifts in lattice units 
at the weakest magnetic field 
in the order of $\rho^0$ (upper) and $\rho^\pm$ (lower). 
The lines correspond to quark masses from the heaviest (circles) to the lightest (triangles).}
\label{emass-rho-m1}
\end{figure}

\begin{figure}
\centerline{\psfig{file=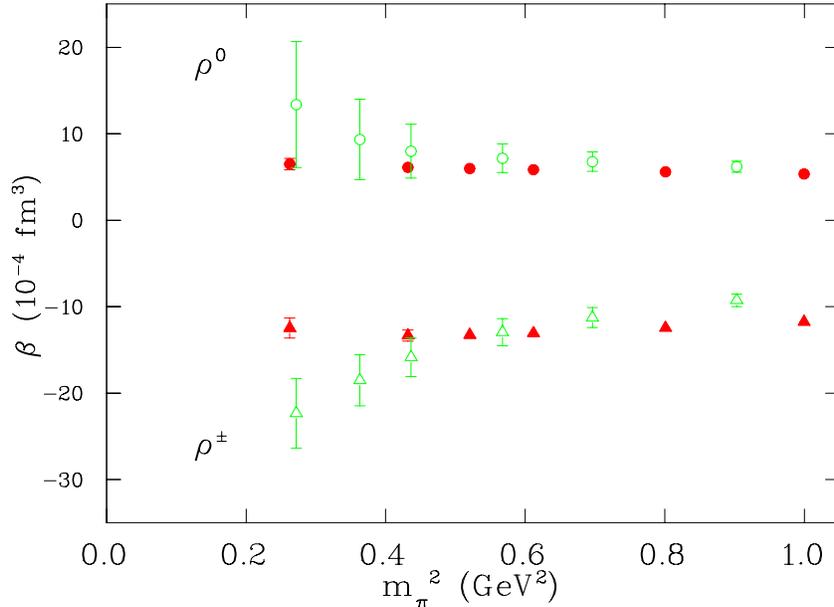,width=11cm,angle=90}}
\caption{Magnetic polarizability for the rho meson states in lattice units.
The solid symbols are the results from the Wilson
action, while the empty symbols are from the clover action.
The Wilson results are obtained from the time window of 9 to 11 for $\rho^0$ and 
14 to 16 for $\rho^\pm$.}
\label{mpol-rho-wc}
\end{figure}

Fig.~\ref{emass-rho-m1} displays the effective mass plot for the rho mesons.
The plateau for $\rho^0$ is not as good as that for $\pi^0$, and we fit in the range 
of 9 to 11. We also fit a bit earlier (range 14 to 16) for $\rho^\pm$ than for 
the pions (16 to 18).
Fig.~\ref{mpol-rho-wc} shows the corresponding magnetic polarizability as a function 
of the pion mass squared.

\begin{figure}
\centerline{\psfig{file=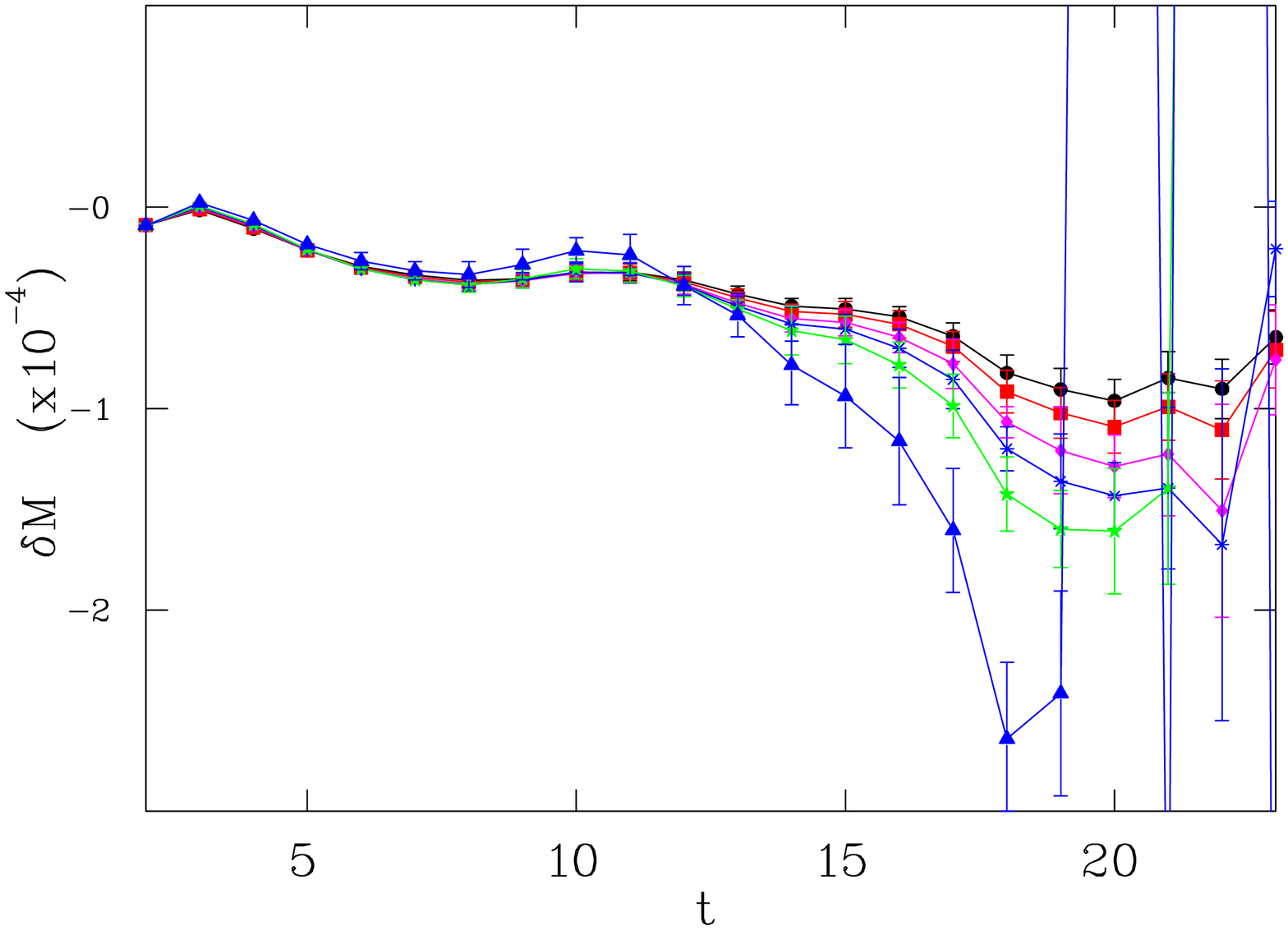,width=7.5cm,angle=90}}
\centerline{\psfig{file=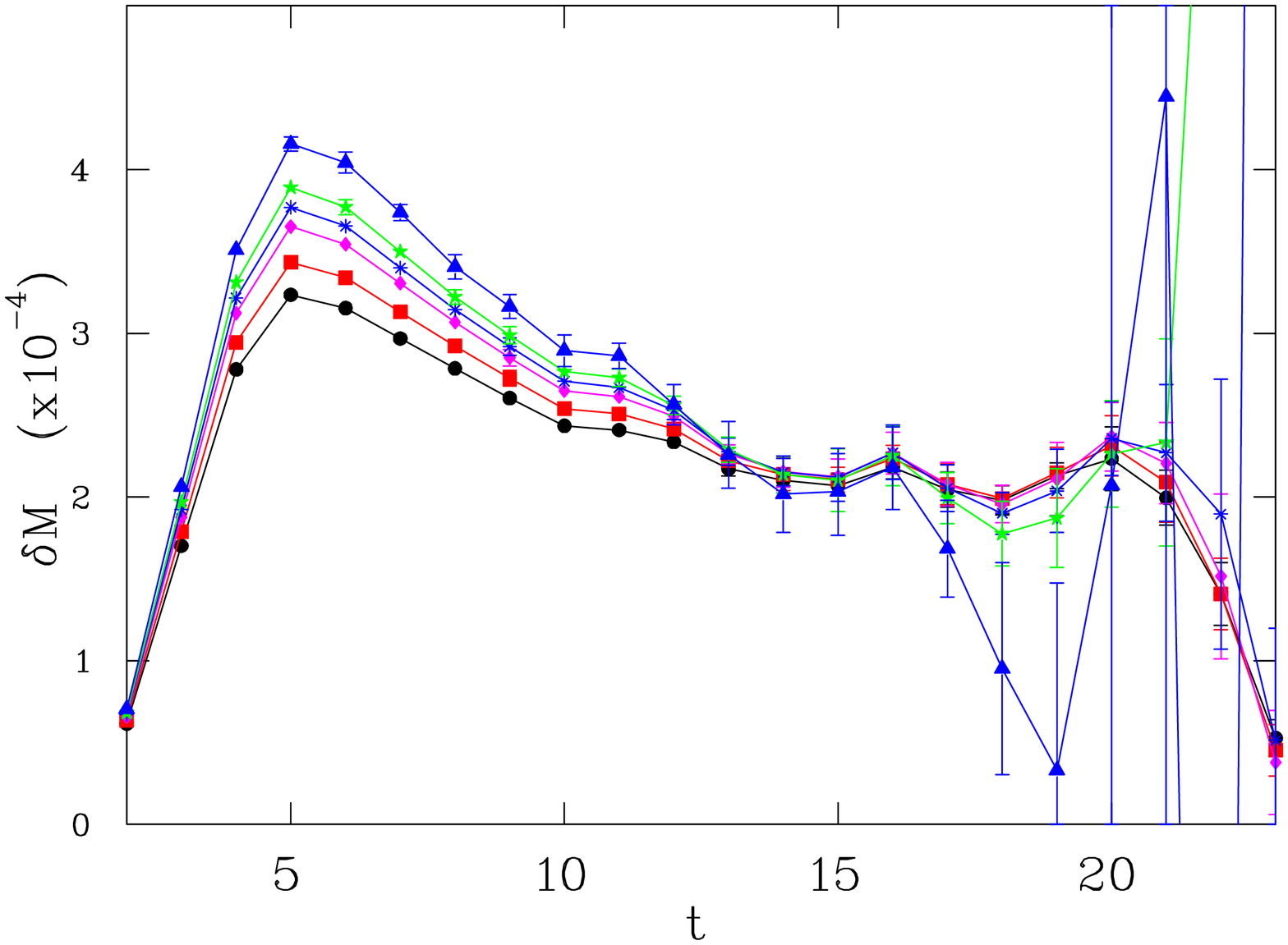,width=7.5cm,angle=90}}
\caption{Effective mass plots for the K* mass shifts in lattice units 
at the weakest magnetic field 
in the order of $K^{*0}$ (upper) and $K^{*\pm}$ (lower). 
The lines correspond to quark masses from the heaviest (circles) to the lightest (triangles).}
\label{emass-Kstar-m1}
\end{figure}

\begin{figure}
\centerline{\psfig{file=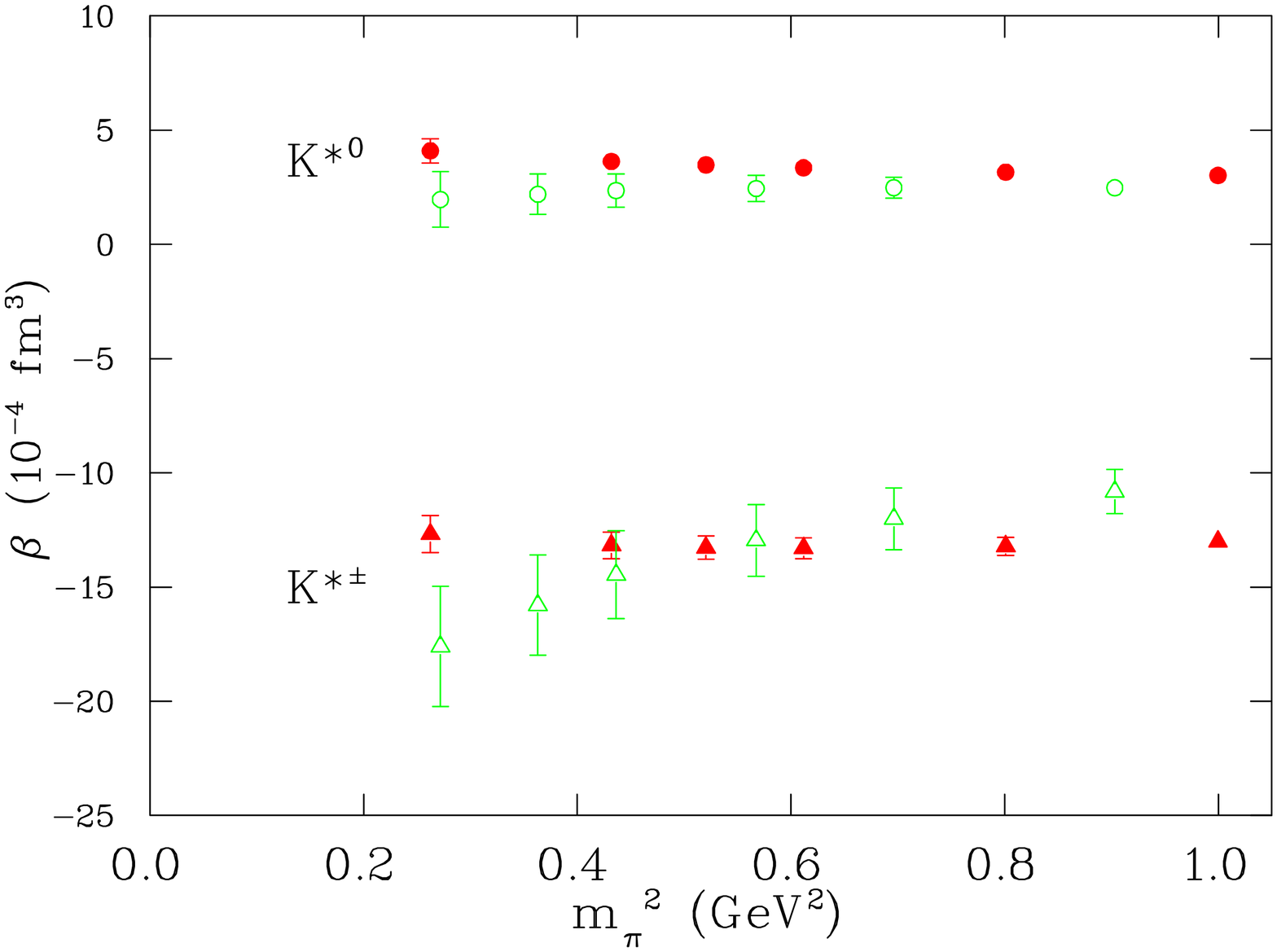,width=11cm,angle=90}}
\caption{Magnetic polarizability for the K* states in physical units.
The solid symbols are the results from the Wilson
action, while the empty symbols are from the clover action.
The Wilson results are obtained from the time window of 14 to 16. }
\label{mpol-Kstar-wc}
\end{figure}

Fig.~\ref{emass-Kstar-m1} displays the effective mass plot for the rho mesons,
and Fig.~\ref{mpol-Kstar-wc} shows the corresponding magnetic polarizability as a function 
of the pion mass squared.

The results of our calculation in the meson sector from the Wilson action are
summarized in Table~\ref{mag-tab-mes}.

Based on the values at the smallest pion mass, one can observe the following
features in the meson sector.
The charged states ($\pi^\pm$, $K^{\pm}$) 
have similar values to each other around $-24$,
which are about twice in magnitude (around $-12$) as the other two pairs 
($\rho^\pm$, $K^{*\pm}$).
The neutral states ($\rho^0$, $K^0$, $K^{*0}$) have 
similar values to each other around $+5$, except $\pi^0$ which 
is about 3 times as large (around $15$).

The most recent measurement of pion polarizability~\cite{Ahrens05} gives 
$(\alpha-\beta)_{\pi^+}=11.6$, which implies $\beta_{\pi^+}$ of about $-6$ 
assuming $(\alpha+\beta)_{\pi^+} \approx 0$ according to leading-order ChPT.
The lattice value of $-24$ has the same sign but is much more negative.
There is no measurement for the rho and K mesons, 
although there are plans to do so~\cite{COMPASS03}.
Other theoretical calculations have produced various values,
about $-2$ for $\pi^\pm$,  and a small but positive value (about $+2$) for $\pi^0$ 
They include ChPT at two-loop order~\cite{chiPT-pi-97}, 
dispersion sum rule (DSR)~\cite{DSR99}, and NJL model~\cite{NJL02}. 
The positive sign for $\pi^0$ is consistent with the lattice value.

In the case of kaon, ChPT predicts $-0.6$ for $K^+$~\cite{Kaon97}.
NJL model predicts $-11$ for $K^+$ and $13$ for $K^0$~\cite{NJL97}.
In the case of vector mesons, the information is extremely limited.
The only calculation is from the QCD string theory~\cite{String99} 
which gives $-0.8$ for rho and $-0.6$ for K*.


\begin{center}
\begin{table*}  
\caption{The calculated magnetic polarizabilities for the selected mesons
as a function of the pion mass from the Wilson action.
The pion mass is in GeV and the magnetic polarizability is
in $10^{-4}$ fm$^3$. The time window from which each polarizability is extracted 
is given in the last column. The errors are statistical.}
 \vspace*{+0.5cm}
\label{mag-tab-mes}
\begin{tabular}{llllllll}
$\kappa$ & 0.1515 & 0.1525 & 0.1535 & 0.1540 & 0.1545 & 0.1555 &  fit range    \\
$ m_\pi$ & 1.000 & 0.895 & 0.782 & 0.721 & 0.657 & 0.512 &     \\
\hline
$\pi^{\pm}$&
  -16.0 $\pm$  0.3&
  -17.8 $\pm$  0.4&
  -19.8 $\pm$  0.5&
  -21.1 $\pm$  0.5&
  -22.5 $\pm$  0.6&
  -26.4 $\pm$  0.8& 16-18  \\
$\pi^{0}$&
   8.4  $\pm$  0.2&
   9.1  $\pm$  0.3&
   10.1 $\pm$  0.4&
   10.8 $\pm$  0.5&
   11.6 $\pm$  0.5&
   14.1 $\pm$  0.8& 16-18 \\
$K^{\pm} $
         & -18.4 $\pm$ 0.4
         & -19.4 $\pm$ 0.4
         & -20.5 $\pm$ 0.5
         & -21.1 $\pm$ 0.5
         & -21.7 $\pm$ 0.6
         & -23.1 $\pm$ 0.6 & 16-18 \\
$K^{0}  $&
   3.7  $\pm$     0.1&
   3.8  $\pm$     0.2&
   4.0  $\pm$     0.2&
   4.1  $\pm$     0.2&
   4.3  $\pm$     0.2&
   4.7  $\pm$     0.3& 16-18  \\
$\rho^\pm$ &
  -11.8 $\pm$      0.3 &
  -12.5 $\pm$      0.3 &
  -13.1 $\pm$      0.4 &
  -13.3 $\pm$      0.5 &
  -13.3 $\pm$      0.6 &
  -12.5 $\pm$      1.2 & 14-16 \\
$\rho^0$ &
   5.3  $\pm$     0.2 &
   5.6  $\pm$     0.2 &
   5.9  $\pm$     0.3 &
   6.0  $\pm$     0.4 &
   6.1  $\pm$     0.4 &
   6.5  $\pm$     0.7 & 9-11 \\
$K^{*\pm} $
         & -13.0 $\pm$ 0.3
         & -13.2 $\pm$ 0.4
         & -13.3 $\pm$ 0.5
         & -13.3 $\pm$ 0.5
         & -13.2 $\pm$ 0.6
         & -12.7 $\pm$ 0.8 & 14-16 \\
$K^{*0}  $&
   3.0  $\pm$     0.2&
   3.2  $\pm$     0.2&
   3.3  $\pm$     0.3&
   3.5  $\pm$     0.3&
   3.6  $\pm$     0.3&
   4.1  $\pm$     0.5& 14-16  \\
\end{tabular}
\end{table*}
\end{center}

\section{Summary and Outlook}

In this work we have presented the results of the first lattice
calculation of hadron magnetic polarizabilities.
Hadron masses were extracted from fits to baryon and meson
two-point correlation functions at six different quark masses and
four different nonzero magnetic fields. 
Magnetic polarizabilities
of hadrons were extracted from the hadron mass shifts induced by the presence of an
external uniform magnetic field. The bulk of our results  
are based on 150 configurations on a $24^4$ lattice at a spacing of about 0.1 fm 
and pion mass down to about 500 MeV, using the standard Wilson actions.
The clover quark action was also used as a check. In both cases, 
clear signals for hadron magnetic polarizabilities are seen and
the calculated results from Wilson
and clover quark actions are generally in good agreement with each other. 
We investigated 30 particles sweeping through the baryon octet and 
decuplet and selected mesons. Most of the polarizabilities have not been measured, 
except for the nucleon and pion, so most of our results are predictions.
Aside from the figures, the results are tabulated in three separate tables.  Below is a summary of the main results of the calculation.

Our value for the proton magnetic polarizability agrees reasonably with
the most recent world average value of about $3$ in units of $10^{-4}$ fm$^3$, 
but suffers from relatively large errors. 
Our value for the neutron magnetic polarizability (about 15 to 20), which has smaller errors,
is much greater than the world average value. 
This large difference between the proton and the neutron on the lattice is an interesting 
result that is worth further study.

In the decuplet sector, the most interesting result is the large and 
negative magnetic polarizability (about $-60$) for the $\Delta^{++}$.
Theoretical explanations, such as those from chiral effective theories, are called for.
The $\Omega^{-}$ value of $-12.4(2)$ is a prediction free of uncertainty
from chiral extrapolations, and can be directly compared with experiment.
A measurement of the $\Omega^{-}$ magnetic polarizability is greatly desired. 

Between the octet and the decuplet, we observe the following pattern.
Positively-charged $p$ and $\Sigma^+$ have relatively small and positive  
values, albeit with large errors, a situation opposite to that for the decuplet 
members $\Delta^+$ and $\Sigma^{*+}$ which have negative and slightly larger values
along with smaller errors.  It is important to increase the statistics to obtain a better 
signal for the $p$ and $\Sigma^+$ on the lattice to confirm this difference.
In addition, the charge-neutral particles have values on the order of $20$, while
the negatively-charged particles on the order of $-20$.

In the meson sector, we confirmed the expected result
that a positively-charged particle has identical magnetic polarizability as its negatively-charged partner ($\pi^\pm$, $K^\pm$, $\rho^\pm$, $K^{*\pm}$).
All charged mesons have negative magnetic polarizabilities, and 
all neutral mesons have positive ones. 
In terms of magnitude, the pseudoscalar mesons have about twice as large values 
as the vector mesons.
Futhermore, the vector mesons have a weaker quark mass dependence than the 
pseudoscalar mesons.

Taken as a whole, our results demonstrate the efficacy of the methods used 
in computing the magnetic polarizabilities on the lattice.
In addition to increasing statistics to the 300 to 500 configurations range,
further studies should focus on assessing the systematic uncertainties. 

First, finite-volume effects must be investigated by changing the lattice size or 
lattice spacing. This is particularly relevant 
considering the fact that the dipole magnetic polarizability is a volume effect 
(in units of fm$^3$). Calculations on a larger lattice $32^4$ are under way.
Another volume is needed to perform a continuum extrapolation.

Next, a chiral extrapolation is needed to provide better comparison with experiment 
and other approaches. To facilitate the chiral extrapolation, on the one hand, 
we need to perform 
calculations at smaller pion masses. This is hard to achieve with the standard Wilson 
quark action due to technical difficulties. An attractive alternate is the so-called 
twisted-mass quark action~\cite{tmqcd} which can push the pion mass down to 
about 250 MeV at a cost of about factor of two.
To further decrease the pion mass, one needs chiral fermions such as the 
overlap~\cite{neu98} which has been used to push the pion mass down to 
about 180 MeV~\cite{uky05}.
The downside of overlap fermions is its cost: it is much more expensive 
to simulate than the standard Wilson fermions. On the other hand, we need 
better ansatz from chiral effective theories to provide a physical basis for the 
extrapolations.

Third, the effects of the quenched approximation should be investigated.
Work is currently under way using the CP-PACS dynamical configurations~\cite{cppacs} based on 
the renormalization-group improved gauge action and mean-field improved clover quark action.
The three lattice sizes in the set are useful for doing a continuum extrapolation.

Finally, as far as the cost of our calculation is concerned, it is equivalent to 11 standard 
mass-spectrum calculations using the same action (5 values of the parameter $\eta$ to 
provide 4 non-zero magnetic fields, 5 to reverse the field, 
plus the zero-field to set the baseline). 
This factor can be reduced to 7 if only two non-zero values of magnetic field are desired.
Reversing the field is well worth the cost: 
the magnetic moments can be extracted from the linear response in the mass shifts 
in the same data set~\cite{Lee05}.

\acknowledgements
This work is supported in part by U.S. Department of Energy
under grant DE-FG02-95ER40907, and by NSF grant 0070836.
W.W. also thanks the Sabbatical Committee
of the College of Arts and Sciences of Baylor University.
The computing resources at NERSC and JLab have been used.


\end{document}